\documentclass[journal]{vgtc}             %
\DeclareGraphicsExtensions{.pdf,.png,.jpg,.jpeg,.eps}

\graphicspath{{figs/}{figures/}{pictures/}{images/}{supp/supp_figs/}{supp/}{./}}

\usepackage{mathptmx}

\usepackage{booktabs}

\onlineid{0}

\vgtccategory{Research}

\usepackage{amsmath}

\usepackage[english]{babel}
\usepackage{stfloats}  %
\usepackage{xspace}
\usepackage{mathtools}
\usepackage[svgnames]{xcolor}

\AtBeginDocument{%
  \captionsetup[subfigure]{font={scriptsize,sf}}%
}

\definecolor{weikaiGreen}{HTML}{196F3D}

\ifx\figurename\undefined \def\figurename{Figure}\fi
\renewcommand{\figurename}{Fig.}

\newcommand{\Sect}[1]{Sec.~\ref{#1}}
\newcommand{\Fig}[1]{Fig.~\ref{#1}}
\newcommand{\Tbl}[1]{Tbl.~\ref{#1}}
\newcommand{\Eqn}[1]{Eqn.~\ref{#1}}

\newcommand{\mode}[1]{\underline{\textsc{#1}}\xspace}

\newcommand{\no}[1]{}
\newcommand{\RNum}[1]{\uppercase\expandafter{\romannumeral #1\relax}}

\newcommand{\cs}{\mathcal{S}}

\ExplSyntaxOn
\cs_new_eq:NN \combined_original_label:n \label
\AtBeginDocument
  {
    \cs_set_eq:NN \combined_original_label:n \label
    \RenewDocumentCommand \label { m }
      {
        \combined_original_label:n {#1}
        \combined_original_label:n {main-#1}
      }
  }
\ExplSyntaxOff

\title{LowPowAR: Power-Constrained Tone Mapping for Augmented Reality}

\let\vgtcauthororcid\authororcid
\renewcommand{\authororcid}[2]{%
  \ifblank{#2}{#1}{\vgtcauthororcid{#1}{#2}}%
}

\author{%
  \authororcid{Weikai Lin}{0000-0003-3537-4857},
  \authororcid{Sheng Zhao}{0009-0008-6532-6999},
  \authororcid{Ian Ross}{ },
  \authororcid{Carl Marshall}{0009-0001-7288-5341},
  \authororcid{Sushant Kondguli}{0000-0002-7295-4626}, and
  \authororcid{Yuhao Zhu}{0000-0002-2802-0578}
}

\authorfooter{
  \item Weikai Lin, Sheng Zhao, and Yuhao Zhu are with the Department of Computer Science, University of Rochester, Rochester, NY, USA.
  E-mail: wlin33@ur.rochester.edu, szhao32@ur.rochester.edu, and yzhu@rochester.edu.
  \item Ian Ross, Carl Marshall, and Sushant Kondguli are with Meta, USA.
  E-mail: iaross@meta.com, csmarshall@meta.com, and sushantkondguli@meta.com.
}

\teaser{
  \centering
  \includegraphics[width=\linewidth]{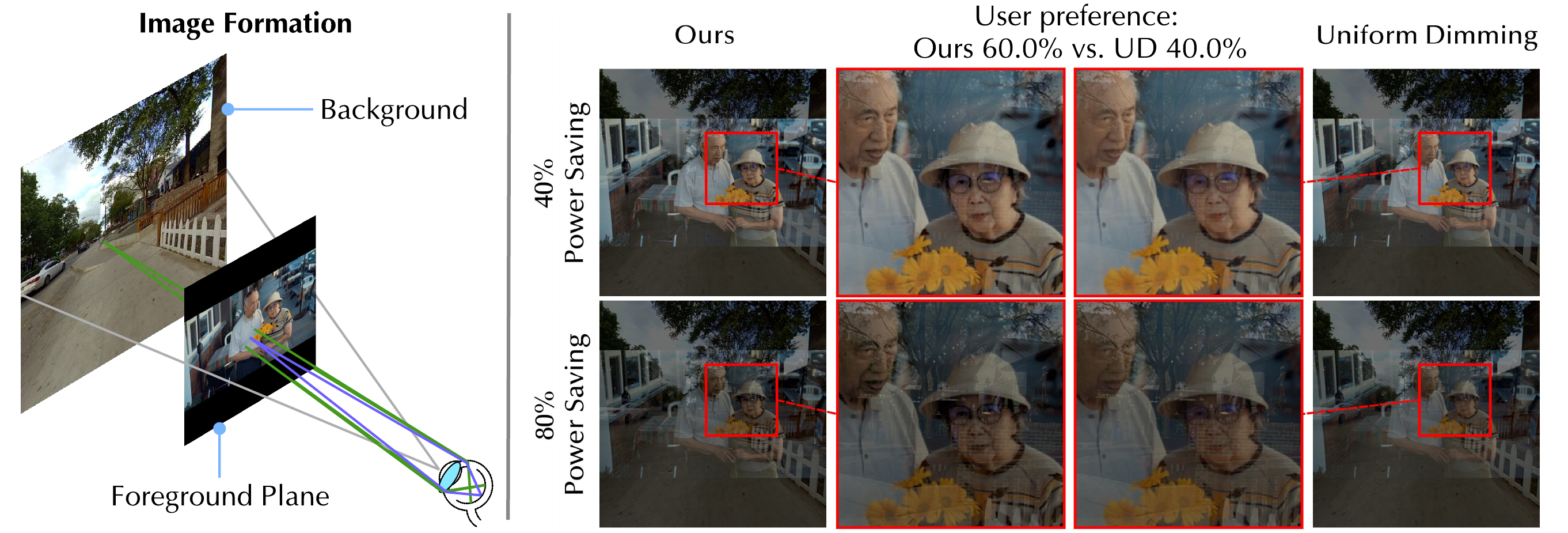}
  \caption{\textbf{Left}: In augmented reality, the perceived image is a mixture of foreground and real-world background, which significantly reduces foreground contrast and readability, especially under aggressive display power saving.
  \textbf{Right}: Our method preserves perceptual quality of the foreground image across different power budgets compare to uniform dimming~\cite{chen2024pea} (and a range of other existing methods; not shown here).
  The foreground-to-background luminance ratio (peak to average) is 30:1.
  The annotated user-preference rate is aggregated over all comparisons in our user study.
  }
  \label{fig:teaser}
}

\abstract{
Everyday-wearable Augmented Reality (AR) glasses must meet strict power limits, making displays a key target for optimization.
We cast display power optimization as a power-constrained tone-mapping problem and propose a human-vision--grounded, learning-based framework that maximizes perceptual quality under a given power budget.
We introduce an optimization-friendly tone-mapping operator (TMO) parameterization along with a progressive optimization strategy to effectively navigate the quality-vs-power landscape.
We distill the iterative optimization into a lightweight feed-forward neural network for real-time deployment.
Subjective experiments show that our method yields better perceptual quality than prior work at the same power budget.
Project page: \url{https://horizon-lab.org/lowpowar/}.
}


\manuscriptnote{}
\begin{document}

\firstsection{Introduction}

\maketitle

\label{sec:intro}

Everyday-wearable Augmented Reality (AR) glasses must operate under an extremely tight power budget, usually around hundreds of milli-Watts.
Display is a significant power consumer and thus, a prime target for power optimizations.
Taking advantage of the fact that the power consumption of emissive displays (such as the $\mu$LEDs widely used in AR glasses) depends on the specific pixel values, several so-called ``display mapping'' methods have been introduced to reduce display power consumption~\cite{chen2024pea, duinkharjav2022color, chen2025modeling, lin2025powergs} by remapping pixel values to alternative ones before sending the image to the display.

We formulate display mapping in optical see-through augmented reality (OST-AR) as a power-constrained \textit{tone mapping} problem, and provide a human-vision--grounded, learning-based framework.
Our framework manipulates image pixel luminance (keeping chromaticity unchanged) to maximize the perceptual quality of the image when viewed against an additive, real-world background while respecting a given display power budget.
Unlike conventional tone mapping, we face three challenges.

First, traditional tone-mapping operator (TMO) design is unconcerned with meeting power targets\footnote{While tone mapping implicitly reduces the power consumption of an image simply by virtue of mapping a high-dynamic-range (HDR) signal to a low-dynamic range (LDR) device, such a mapping does not respect a given power budget apart from capping the peak power usage limited by the maximum luminance of the LDR display.}.
It remains open how to design a TMO to effectively navigate the power-vs-quality landscape.
Recent work uses either a global dimming factor~\cite{chen2024pea} or a per-pixel dimming map~\cite{le2023energy} to control the power.
The former is too restrictive while the latter introduces a large degrees of freedom, complicating optimization.
We propose a learning-friendly TMO parameterization that allows us to flexibly trade perceptual quality for power saving (\Sect{sec:tonemapper}).

Second, existing perceptual metrics~\cite{mantiuk2024colorvideovdp, mantiuk2021fovvideovdp, cogalan2025milo, zhang2018unreasonable}) are ineffective \textit{loss functions}.
This is because they are fit to a specific set of artifacts, such as blur or noise.
Learning the TMOs under an aggressive power constraint and additive background using these metrics, however, will necessarily explore stimuli in previously unseen spaces, encountering ``adversarial stimuli'' that look subjectively poor but have a low loss.
As a result, gradients become unreliable, posing challenges to optimization.
We propose a novel progressive optimization strategy to side-step this issue (\Sect{sec:loss}).

Finally, prior TMOs either are obtained through heavy offline processing (when real-time processing is not the goal) or have to resort to relatively simple (global) TMOs to be applied in real-time~\cite{tariq2023perceptually}.
We address this by distilling the iterative, offline optimization into a lightweight neural-network--based pipeline.
This enables real-time inference while preserving quality (\Sect{sec:realtime}).

In summary, our primary contributions are:

\begin{itemize}
    \item We formulate display power optimization as a power-constrained tone mapping problem in the context of AR and provide a perceptually-aware optimization framework that learns to tone map an image under a given power budget while maximizing perceptual quality.
    \item We show how to distill the optimization procedure into a light-weight algorithm for real-time deployment.
    \item Through extensive user study and ablation studies, we show that our method out-performs a diverse set of prior methods.
\end{itemize}

\section{Related Work}
\label{sec:related}

\subsection{Perceptual Tone Mapping}
\label{sec:related:tm}

Tone mapping techniques from the very beginning are driven by human vision and viewing condition~\cite{tumblin2002tone, larson1997visibility}.
Along with recent advances in perceptual quality metrics~\cite{mantiuk2021fovvideovdp, mantiuk2024colorvideovdp, ashraf2022suprathreshold}, recent tone mapping work has explicitly incorporated models of human vision to optimize the TMOs~\cite{eilertsen2015real, mantiuk2008display, tariq2023perceptually}.
We refer interested readers to recent texts for a review \cite{eilertsen2017comparative, reinhard2010high, mantiuk2015high} .

Wanat and Mantiuk~\cite{wanat2014simulating} consider tone mapping to extremely dim displays where vision is (partially) modulated by scotopic vision, but it does not guarantee a power budget.
In addition, in most AR settings even if the display is dimmed, user's vision is still mostly modulated by photopic vision.

Very few TMOs enforce display power budgets or account for the additive real-world background.
Learning the TMO under a power constraint is a difficult, multi-objective optimization problem.
A critical question is how to parameterize a training-friendly TMO that also allows to freely explore the quality-vs-power space.
Recent display mapping work uses either overly flexible parameterizations (e.g., per-pixel scaling)~\cite{le2023energy} or overly constrained parameterizations (e.g., a global dimming factor)~\cite{chen2024pea};
neither gives desirable results.
Our contribution is a learning-friendly TMO parameterization.

We also show that many excellent perceptual metrics, such as ColorVideoVDP~\cite{mantiuk2024colorvideovdp} and MILO~\cite{cogalan2025milo}, are sub-optimal when used as a loss functions to train the TMOs, especially in low-power AR scenarios, where there is a significant appearance difference between the reference image and the desired image.
We propose an optimization strategy that meets the power budget while retaining high perceptual quality.

\subsection{Display Power Optimizations}
\label{sec:related:dp}

Display has long been a prime target for power reduction, going back to the early days of smartphone displays~\cite{kerofsky2006brightness, chang2004dls, dash2021much, halpern2016mobile}.
It has been even more important to AR/VR systems, where the display power is even higher to due the quest for higher pixel density, temporal frame, and peak brightness~\cite{lin2025powergs, chen2025modeling, chen2024pea, chen2026mlpea, duinkharjav2022color, leng2019energy}.

Chen et al.~\cite{chen2025modeling} propose a gaze-free, color modulation technique exploiting chromatic-adaptation kinetics. 
Chen et al.~\cite{chen2024pea} show that under a fixed power budget, uniform dimming, which is a gaze-free and luminance-only method, out-performs a range of gaze-contingent techniques, such as color modulation based on discrimination contours~\cite{duinkharjav2022color}.
Our tone mapping method is also gaze-free, modulates only luminance, and out-performs uniform dimming.
Prior OST-AR rendering methods enhance color contrast to improve visibility~\cite{zhang2021contrast}, but do not formulate tone mapping as perceptual optimization under an explicit display-power saving target.
Complementary OST-AR studies characterize brightness, transparency, contrast, and display artifacts under additive viewing conditions~\cite{murdoch2020brightness,zhang2021perceived,herbeck2024transparency,chapiro2024ar,kim2025supra}.
These studies characterize perception and artifacts rather than optimize displayed content for a prescribed power budget.

In general, our work falls under the umbrella of perceptual rendering driven by models of human vision~\cite{tursun2019luminance, walton2021beyond, lin2025metasapiens}.
Tursun et al.~\cite{tursun2019luminance} use a contrast-based loss similar to that used in ColorVideoVDP~\cite{mantiuk2024colorvideovdp}, which this work relies on.
Surace et al.~\cite{10.2312:sr.20251183} dynamically apply a global dimming factor to reduce display power under a contrast loss constraint.

\section{Problem Formulation}
\label{sec:problem}

Given an input image $I$, a target display power budget $P_t$, and an image $L_{\text{bg}}$ representing the real-world background,  $f(\cdot)$ is a display mapping function that maps the image $I$ to an alternative image $I_{mapped}$.
The final perceived image $L_{\text{perceived}}$ is then given by:
\begin{subequations}
    \label{eq:ar_composite}
    \begin{align}
        f: I \mapsto I_{mapped}, \label{eq:ar_composite_1}\\
        L_{\text{perceived}} = \mathcal{D}(I_{mapped}) + L_{\text{bg}} \approx \mathcal{D}(I_{mapped}) + \bar{L}_{\text{bg}}, \label{eq:ar_composite_2}
    \end{align}
\end{subequations}
where $\mathcal{D}(\cdot)$ is the display model that maps the pixel values in $I_{mapped}$ to absolute luminance measures~\cite{mantiuk2024colorvideovdp, mantiuk2008display}.
We use the well-known gamma-offset-gain (GOG) model (see Supplementary Material \ref{sec:supp:display_model} for details).

Rather than using the precise background image, we represent the background as a uniform achromatic field using the mean luminance of the background image $\bar{L}_{\text{bg}}$.
We do so for two reasons.
First, prior work has shown that using a uniform background better reflects perceived quality of additive content than using the precise background~\cite{chapiro2024ar}.
Second, this design significantly reduces hardware requirements.
Capturing the precise background would require a near-zero-latency camera with geometric warping to align foreground and background content.
In contrast, using the mean ambient luminance requires only a simple ambient luminance meter~\cite{metarayban2025, magicleap2022}.

Our optimization objective is to minimize the perceptual loss $\mathcal{L}_{\text{perceptual}}$ between the perceived image and the reference image while satisfying a given power constraint:
\begin{equation}
    \min_{f} \mathcal{L}_{\text{perceptual}}(L_{\text{perceived}}, \mathcal{D}(I)) \quad \text{s.t.} \quad \mathcal{P}(f(I)) \leq P_t,
    \label{eq:objective}
\end{equation}
where $I$ is the reference image (i.e., without any tone mapping and the additive background image),
$\mathcal{P}(\cdot)$ computes the display power given an image.
We use an OLED display power model widely used in prior work where a pixel's power consumption is modeled as weighted sum of linear sRGB values~\cite{duinkharjav2022color}.

\section{Method}
\label{sec:method}

\begin{figure*}[t]
    \centering
    \includegraphics[width=\linewidth, alt={Overview of the offline progressive optimization and its distillation into a real-time network. The pipeline decomposes the foreground into base and detail layers, applies learned tone mappings, simulates additive OST-AR compositing, and optimizes perceptual quality subject to a power target.}]{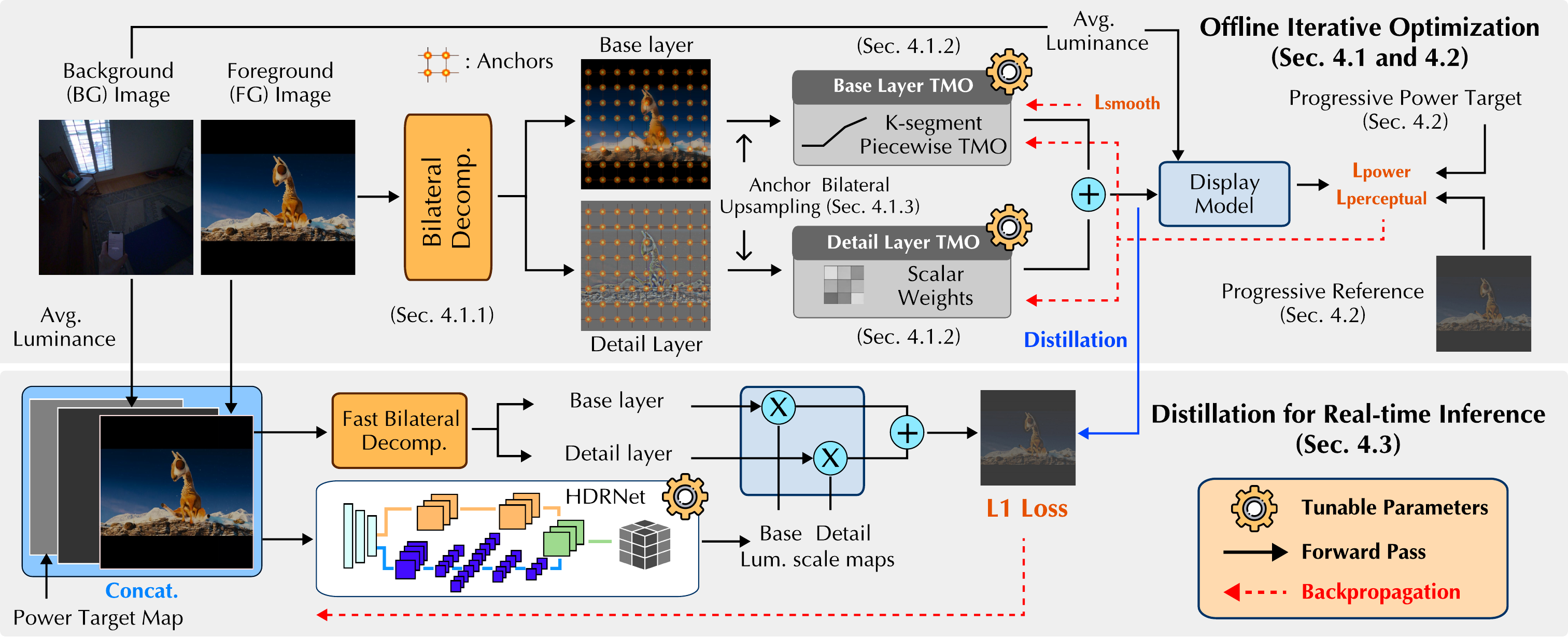}
    \caption{\textbf{Framework overview.}
    \textbf{Top:} Our offline iterative optimization decomposes the foreground image into base and detail layers via bilateral filtering.
    We optimize per-pixel piecewise-linear tone curves (K-segment) for the base layer and per-pixel weighting for the detail layer.
    A display simulation model combines the mapped image with the average background luminance, and we minimize a combination of perceptual, power, and smoothness losses.
    The progressive reference is updated iteratively to keep the perceptual metric in its reliable operating range.
    \textbf{Bottom:} We distill the optimization results into a real-time pipeline that directly predicts luminance scaling maps during inference, conditioned on the target power budget and background luminance.}
    \label{fig:overview}
\end{figure*}

We present a perception-driven display mapping framework for power-constrained AR displays.
We begin by describing a trainable parameterization of the TMO (\Sect{sec:tonemapper}), followed by an offline solver using progressive optimization (\Sect{sec:loss}).
We then distill the offline optimization into a lightweight neural network for real-time inference (\Sect{sec:realtime}).
\Fig{fig:overview} illustrates our framework.

\subsection{Trainable TMO Parameterization}
\label{sec:tonemapper}

Our TMO is based on bilateral decomposition~\cite{tomasi1998bilateral} commonly used in HDR tone mapping~\cite{durand2002fast} that decomposes the image into an edge-preserving, low-frequency base layer and a high-frequency detail layer.
In Retinex terms~\cite{land1971lightness}, the base layer represents the illumination of the scene and, thus, carries most of the power of the original image signal, so scaling down the base layer directly translates to power reduction.

At the same time, perceptual quality can be controlled by properly manipulating the base and detail layers.
We apply local tone mapping to the base layer to control the local adaptation levels, exploiting the fact that human perception is (at least partially) tuned to local contrasts.
We also apply local tone mapping to the detail layer, which could potentially boost details near contrast thresholds while suppressing noises~\cite{eilertsen2015real}.
We now give the mathematical formulation of the TMO parameterization.

\subsubsection{Base-Detail Decomposition}
\label{sec:tonemapper:bd}

We decompose the input image $I$ into a base layer $I_{\text{base}}$ capturing edge-stopping, averaged local illumination and a detail layer $I_{\text{detail}}$ preserving high-frequency signals:
\begin{equation}
    \label{eq:decomp}
    I_{\text{base}} = \mathcal{B}(I; \sigma_s, \sigma_r), \quad
    I_{\text{detail}} = I - I_{\text{base}},
\end{equation}
where $\mathcal{B}(\cdot)$ represents bilateral filtering, $\sigma_s$ and $\sigma_r$ control the spatial and range kernels of the bilateral filter, respectively.

\subsubsection{Local TMO Parameterization}
\label{sec:tonemapper:tmo}

To give sufficient freedom in exploring the quality-vs-power trade-offs, we define a sparse grid of \emph{anchors} $\mathcal{A}$ at resolution $H_a \times W_a$, where $H_a = H/\Delta$ and $W_a = W/\Delta$ for an image with resolution $H \times W$ and anchor stride $\Delta$.
Each anchor receives a separately trainable local TMO for the base layer and the detail layer.
The TMO parameters are then upsampled (see below) from the anchors to every pixel $(x, y)$ in the image.

The local TMO in the base layer for a pixel $(x, y)$ is a $K$-segment, monotonic, piecewise-linear function~\cite{mantiuk2008display, eilertsen2015real}\footnote{Many classical (piecewise-linear) TMOs operate in luminance-log space inspired by the Weber-Fechner law~\cite{fechner1860elemente}. Our trainable TMOs operate in the luminance-linear space, because (approximate) log transformation could be easily learned if effective.}, parameterized by the slope of each of segment: $\theta_{x,y} = \{\delta_{x, y}^0, \delta_{x, y}^1, \ldots, \delta_{x, y}^{K-1}\}$, where $\delta_k \geq 0$ represents the slope of the $k$-th segment (assuming the intercept of the first segment is 0).
For a pixel with a normalized luminance $I_{base}(x, y) \in [0, 1]$ in the base layer, the output $I_{base}'(x, y)$ is computed by:

\begin{equation}
    \label{eq:base}
    I_{base}'(x, y; \theta_{x, y}) = \sum_{k=0}^{K-1} \delta_{x, y}^k \cdot \min(\max(\frac{I_{base}(x, y) - k/K}{1/K}, 0), 1).
\end{equation}

The monotonic piecewise-linear function provides a high degree of freedom for exploring power-vs-quality trade-off (over other common TMOs such as the gamma or sigmoid functions) while ensuring that brighter pixels remain brighter after mapping.
Since the original image and the mapped image are to be displayed on the same display, the TMO $I_{base}(x, y) \mapsto I'_{base}(x, y)$ operate within the normalized [0, 1] range, where 1 corresponds to the peak nit of the display.
The normalized pixel values will be mapped to the actual luminance using the display model.

The local TMO for each pixel $(x, y)$ in the detail layer is simply a scalar weight $w_{x, y}$ that allows for local contrast modulation, which can either improve visibility of features near the contrast thresholds or suppress noise~\cite{eilertsen2015real}.
The output is given by:
\begin{align}
    \label{eq:detail}
    I_{detail}'(x, y) = w_{x, y} \cdot I_{detail}(x, y).
\end{align}

We observe that a more expressive TMO (e.g., a piecewise function) in the detail layer is unnecessary because: 1) it may produce unnatural local contrast changes and 2) the detail layer has a minor impact on display power, so it does not require as much flexibility in balancing power consumption against perceptual quality.

\subsubsection{Guided Upsampling and Reconstruction}
\label{sec:tonemapper:upsample}

To obtain full-resolution TMOs for every pixel, we upsample the TMO parameters from the sparse anchor grid $\mathcal{A}$ using joint bilateral upsampling~\cite{kopf2007joint, chen2016bilateral} with the input image $I$ as the guide.
This allows for edge-preserving parameter upsampling that respects image boundaries.
For each pixel $p$ in the full-resolution image, we compute its base-layer TMO parameters $\theta_p$ by interpolating from its nearby anchors:
\begin{equation}
    \label{eq:upsample}
    \theta_{p} = \frac{\sum_{q \in \mathcal{N}_{p}} B(p, q) \cdot \theta_{q}}{\sum_{q \in \mathcal{N}_{p}} B(p, q)}, \quad
    B(p, q) = G_s(\|p_\downarrow - q\|) \cdot G_r(\|I_p - I_{q_\uparrow}\|),
\end{equation}
where $\mathcal{N}_{p} \subset \mathcal{A}$ denotes the set of neighboring anchors around pixel $p$ used for upsampling, $q$ is the anchor coordinate at anchor resolution, $p_\downarrow$ is pixel $p$ downsampled to anchor resolution for computing spatial distance, $I_{q_\uparrow}$ is the pixel value at the nearest-neighbor upsampled location of anchor $q$, $B(p, q)$ is the bilateral weight combining spatial kernel $G_s$ and range kernel $G_r$ with standard deviations $\sigma_s$ and $\sigma_r$, respectively, and the denominator serves as a normalization factor.
The tunable parameters include the number of neighboring anchors $|\mathcal{N}_{p}|$ and the kernel widths $\sigma_s$ and $\sigma_r$.
The detail layer weights $w_{x,y}$ are upsampled similarly.

The output image is reconstructed by applying the (upsampled)  display mapping to each layer and combining the results:
\begin{equation}
    \label{eq:recomp}
    I_{mapped}(x, y) = I_{base}'(x, y) + I_{detail}'(x, y).
\end{equation}

Cascading \Eqn{eq:decomp}, \Eqn{eq:base}, \Eqn{eq:detail}, \Eqn{eq:upsample}, and \Eqn{eq:recomp} forms the display mapping function $f(\cdot)$ in \Eqn{eq:ar_composite_1}.

\subsection{Progressive Offline Optimization}
\label{sec:loss}

We describe an optimization strategy to obtain the base and detail TMOs and, thereby, solve the general optimization problem in \Eqn{eq:objective}.
This optimization is not meant for real-time execution (it takes about 2 minutes on an RTX 5090), but acts as a ``teacher'' model that will later be distilled into a lightweight pipeline (\Sect{sec:realtime}).

\subsubsection{Power Loss}
\label{sec:loss:power}

The power constraint is turned into an additional power penalty term in the loss function.
Specifically, we use a conditional power loss that penalizes power consumption only when exceeding the target:
\begin{equation}
    \label{eq:powerloss}
    \mathcal{L}_{\text{power}} =
    \begin{cases}
        0 & \text{if } \mathcal{P}(I_{\text{mapped}}) \leq P_t \\
        \lambda_p \cdot \mathcal{P}(I_{\text{mapped}}) & \text{otherwise}
    \end{cases},
\end{equation}
where $\lambda_p$ is a large weight.

\subsubsection{Selective Global Smoothing.}
\label{sec:loss:smoothing}

Local tone mapping, especially in the base layer, is prone to produce spatial non-uniformity, which is particularly noticeable in content with large uniform regions such as reading texts on a webpage.
To address this, we regularize the TMOs of the anchors in the base layer toward their global mean:
\begin{equation}
    \label{eq:global}
    \mathcal{L}_{\text{smooth}} = \lambda_s \sum_{i,j} \|\theta_{i,j}^{\text{base}} - \bar{\theta}^{\text{base}}\|^2,
\end{equation}
where $\bar{\theta}^{\text{base}} = \frac{1}{|\mathcal{A}|} \sum_{i,j} \theta_{i,j}^{\text{base}}$ is the global mean of all the anchors and $\lambda_s$ represents the strength of smoothing.
We apply smoothing only to the base layer because non-uniform luminance primarily arises from it.
The detail layer, which controls local texture contrast, benefits from spatial variation and is left unconstrained.

Globally smoothing the TMOs has the additional benefit of optimizing the global contrast, and the smoothing term effectively blends the effect of local and global tone mapping, a technique commonly used in prior tone mapping work~\cite{eilertsen2015real, tariq2023perceptually, hasinoff2016burst}.

\subsubsection{Progressive Optimization.}
\label{sec:loss:opt}

We use ColorVideoVDP~\cite{mantiuk2024colorvideovdp}, a differentiable, contrast-based quality metric, as our perceptual loss $\mathcal{L}_{\text{perceptual}}$.
ColorVideoVDP forms the perceptual loss mainly by modeling the contrast perception of the human visual system, which is shown to be critical for evaluating subjective experience in tone mapping~\cite{eilertsen2015real, mantiuk2008display, tariq2023perceptually} and in AR additive displays~\cite{kim2025supra}.
Critically, ColorVideoVDP models the luminance dependence of the contrast sensitivity function (CSF), which allows us to account for display characteristics and background luminance levels.

While ColorVideoVDP is an effective perceptual metric, we find it to be ineffective as a \textit{loss function} to train TMOs directly.
This is because ColorVideoVDP parameters are fit to a specific set of distorted stimuli, such as blur or noise, that do not usually appear during our training.
Our TMO training will necessarily explore stimuli in previously unseen spaces, encountering ``adversarial examples'' that look objectionable but score highly with the metric (see Supplementary Material \ref{sec:supp_perceptual} for concrete examples).
Adversarial examples occur for two reasons: 1) power reduction lowers the overall luminance, whereas ColorVideoVDP considers only distortions that do not alter luminance much, and 2) the additive background shifts the perceived image significantly away from the original content.
The gradients become less meaningful under adversarial examples, deteriorating the optimization quality.

To address this, we propose a progressive optimization strategy, where the reference/ground-truth images are dynamically updated during training.
The key idea is to always compare images with only slight differences.
Rather than optimizing directly to the target power $P_t$ (see \Eqn{eq:powerloss}), we decompose the optimization into $N$ stages with intermediate power targets $P_t^1, P_t^2, \ldots, P_t^N$, where the power target at stage $k$ is $P_t^k = 100\% - k \cdot \Delta P$ ($\Delta P = 10\%$ in our experiment).
At stage $k$, we use the optimized image from stage $k{-}1$ to form the reference, with $I_{\text{mapped}}^{(0)} = I$ being the original image at 100\% power:

\begin{equation}
    L_{\text{ref}}^{(k)} = \mathcal{D}(I_{\text{mapped}}^{(k-1)}) + \bar{L}_{\text{bg}},
\end{equation}

\noindent where $L_{\text{ref}}^{(k)}$ is the reference image at stage $k \in [1, N]$.

The progressive strategy is effective because it decomposes a difficult optimization into multiple easier sub-problems.
Directly optimizing toward a low-power target using the original, full-power image as reference creates a large appearance gap, resulting in many possible optimization paths, some of which might contain adversarial examples with unreliable gradients.
In contrast, by using a slightly higher-power image as the reference in each stage, the reference and target remain close, reducing the chance of encountering adversarial examples and yielding more reliable gradients.

\subsubsection{Putting It Together}

The optimization problem in \Eqn{eq:objective} is re-formulated as:

\begin{equation}
    \label{eq:finalloss}
    \min_{\mathbf{\theta^k}, \mathbf{w^k}} \mathcal{L}_{\text{perceptual}}(\mathcal{D}(f(I; \mathbf{\theta^k}, \mathbf{w^k}) + \bar{L}_{bg}), L_{\text{ref}}^{(k)}) + \mathcal{L}_{\text{power}} + \mathcal{L}_{\text{smooth}},
\end{equation}

\noindent where $\mathcal{L}_{\text{perceptual}}$ is the perceptual loss, $\mathbf{\theta^k}$ and $\mathbf{w^k}$ are the base and detail TMOs of the anchors $\mathcal{A}$, respectively, in the $k$-th stage, and $L_{\text{ref}}^{(k)}$ is the progressive reference at stage $k$.
At each stage, we initialize with the previous stage's result and perform $T$ iterations of gradient descent to minimize the loss.
The complete optimization runs for $N \times T$ iterations in total.

\subsection{Distillation for Real-time Inference}
\label{sec:realtime}
While iterative optimization produces high-quality results, it is too slow for real-time applications.
We distill the optimization into a lightweight neural network that directly predicts the TMOs.
We generate the training set across diverse images, power budgets, and background conditions using the algorithm in \Sect{sec:loss}.

Our inference pipeline is shown in \Fig{fig:overview}.
The network takes three inputs: the foreground image $I$, a target power $P_t$, and the average background RGB value $\bar{I}_{\text{bg}}$.
We first perform fast bilateral decomposition to separate the image into base and detail layers~\cite{paris2006fast}.
We keep this decomposition outside the network, as we found that high-frequency detail extraction through bilateral filtering is difficult to learn.

Both $P_t$ and $\bar{I}_{\text{bg}}$ are broadcast to match the resolution of $I$, and then concatenated with $I$ before entering the network.
The network, in one invocation, predicts a per-pixel luminance scaling map for the base and detail layer separately.
The scaling maps are then applied to the two layers, which are then combined together to generate the output image.
We predict the scaling maps separately for the two layers to match how they are dealt with in the offline optimization procedure (\Sect{sec:loss}):
the base layer captures local (averaged) luminance and carries most of the image power while the detail layer encodes local contrast, so they require different treatments.

We use HDRNet~\cite{gharbi2017deep} as the backbone of our neural network for three primary reasons.
First, HDRNet is specifically designed to be computationally efficient, including on mobile hardware.
Second, our goal is to adjust each pixel's luminance, which can be expressed as a constrained affine transformation of its RGB value---the type of transformation HDRNet is built to predict.
Third, HDRNet uses a bilateral grid to steer the upsampling process~\cite{chen2016bilateral}, which is a generalization of joint bilateral upsampling ~\cite{kopf2007joint} that we use in offline iterative optimization.

\section{Experiments}
\label{sec:exp}
\subsection{Experiment Setup}
\label{sec:exp:setup}

\paragraph{Power Modelling.}
We estimate display power using an OLED power model from prior work~\cite{duinkharjav2022color}, which has been validated against real hardware and is widely adopted in display power research~\cite{chen2024pea, lin2025powergs, chen2025modeling}.

\paragraph{AR Viewing Conditions.}
Like AR-DAVID~\cite{chapiro2024ar}, we quantify a specific AR viewing conditions using the \emph{foreground-to-background luminance ratio} (FG:BG), i.e., the ratio of foreground peak luminance to background average luminance.
This ratio determines the peak contrast of the foreground content when adapted to the real-world scene.
To establish a representative range for this ratio, we survey typical luminance values in AR scenarios.
Modern AR displays achieve peak luminance from 700 to 5,000 nits~\cite{xreal2026glasses, magicleap2022, metarayban2025}.
Ambient luminance varies by environment, typically ranging from 30 to 10,000 nits~\cite{hilmers2022quantification}.
Thus, we use FG:BG ratios ranging from 3:1 to 30:1, consistent with the range used in prior AR perception studies~\cite{chapiro2024ar}.

\paragraph{Physical Setup.}

\begin{figure*}[t]
    \centering
    \includegraphics[width=\linewidth, alt={Left, a schematic in which a beam splitter reflects foreground monitor light and transmits background monitor light toward the viewer. Right, a participant views three side-by-side stimuli through the beam splitter while seated at a chin rest.}]{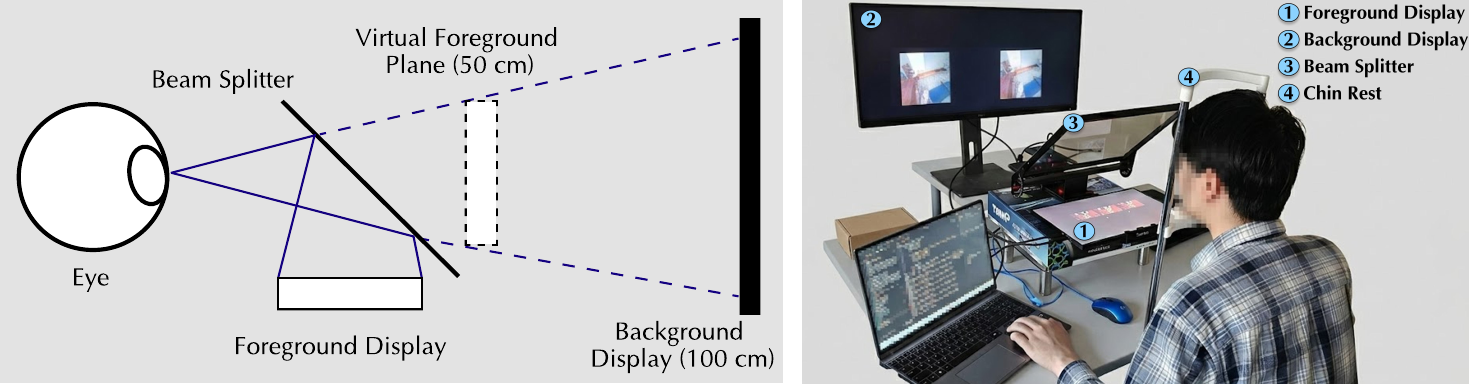}
    \caption{\textbf{Experiment setup.}
    \textbf{Left}: We simulate an optical see-through AR display using two monitors and a beam splitter. The foreground content is displayed on a 1,200-nit monitor and reflected by a 70:30 beam splitter, yielding an effective peak luminance of 360 nits. The background is displayed on a monitor behind the beam splitter.
    \textbf{Right}: Physical setup for the user study. 
    A participant views three side-by-side images through the beam splitter: the original reference at the center, with tone-mapped results from our method and a baseline on each side. The participant selects the variant that more closely matches the reference using the keyboard. A chin rest controls the viewing distance.}
    \label{fig:experiment_setup}
\end{figure*}

\Fig{fig:experiment_setup} illustrates our setup with two monitors and a beam splitter.
The foreground content is displayed on a 16-inch 4K  monitor\footnote{Newsoul 4K Portable Monitor, 3840$\times$2400 resolution, 1200-nit peak luminance, 2000:1 contrast ratio.} with a peak luminance of 1,200 nits.
The background is displayed on a 34-inch ultra-wide monitor\footnote{AOC U34P2 34-inch professional-grade ultra-wide monitor, 3440$\times$1440 resolution, 350-nit peak luminance.} with a peak luminance of 350 nits.
We use a beam splitter with a 70:30 transmittance-to-reflectance ratio.
Foreground content is presented on the first display reflected by the beam splitter, yielding a maximum luminance of 360 nits, similar to prior AR perception studies~\cite{chapiro2024ar}.
The background is displayed on the second monitor placed behind the beam splitter.
The viewing distance of the foreground is 50 cm, and the background distance is
100 cm.
Distances are controlled using a chin rest.
We calibrate the monitor brightness and the beam splitter transmittance/reflectance using an SM208 luminance meter.
\paragraph{Dataset.}
Our evaluation requires paired foreground (displayed content) and background (real-world scene) images.
We construct two datasets, resizing and padding all images to $800 \times 800$, similar to existing AR devices~\cite{metarayban2025}:
\begin{itemize}
    \item \textbf{Evaluation set:} 17 foreground images from AR-DAVID~\cite{chapiro2024ar} and XR-DAVID~\cite{chapiro2024xr} paired with 11 backgrounds: 5 egocentric captures from the Aria dataset~\cite{pan2023aria}, 3 outdoor scenes from PEA-PODs~\cite{chen2024pea}, and 3 synthetic patterns (dead leaves, uniform, and pink noise) from AR-DAVID~\cite{chapiro2024ar}.
    This yields 187 foreground-background pairs covering diverse real-world and controlled conditions.
    We show all images in the Supplementary Material~\ref{sec:supp:dataset}.
    \item \textbf{Training set:} 200 foreground images: 100 sampled from LAION-POP~\cite{laion_pop}, a dataset of popular web images, and 100 sampled from RICO~\cite{deka2017rico}, a mobile UI screenshot dataset. 
    All are paired with uniform backgrounds at three FG:BG ratios (3:1, 10:1, and 30:1), as our method requires only uniform backgrounds as inputs (\Sect{sec:exp:ablation_bg}).
    This training set is used for distillation (\Sect{sec:realtime}).
\end{itemize}

Critically, notice that the evaluation set and the training set come from completely different datasets.
The evaluation set consists of exclusively prior scenes used in XR studies, whereas the training set consists exclusively of non-XR generic scenes.
This allows us to demonstrate the generalizability of our framework.

\paragraph{Iterative Optimization Setup.}
\label{sec:exp:iter}
To parameterize the TMOs (\Sect{sec:tonemapper}), the base layer uses a piecewise linear function with $K=10$ segments and an anchor stride $\Delta = 128$.
The detail layer shares the same anchor configuration, with one scaling factor associated with each anchor.
For the loss function (\Sect{sec:loss}), we use $\lambda_p = 10{,}000$ for controlling the power loss and $\lambda_s = 10$ for smoothness regularization.
In progressive training (\Sect{sec:loss}), we use a power step $\Delta P = 10\%$ and run $T = 300$ iterations per stage.
The learning rate for optimizing the TMO parameters is set to $10^{-2}$.
For bilateral filtering we use ($\sigma_s$, $\sigma_r$) = (15, 0.06).
For bilateral upsampling we use ($\sigma_s$, $\sigma_r$) = (0.5, 0.1) and $|\mathcal{N}_{p}|$ includes 5 $\times$ 5 nearby anchors
($\sigma_s$ = 0.5 here means 64 pixels as anchor size is 128).

\begin{figure*}[t]
    \centering
    \includegraphics[width=0.95\linewidth, alt={Two bar charts of LowPowAR win rates against NoProg, uniform dimming, EAI, and PCSR. Every bar is above the 50 percent chance line under both tested viewing conditions.}]{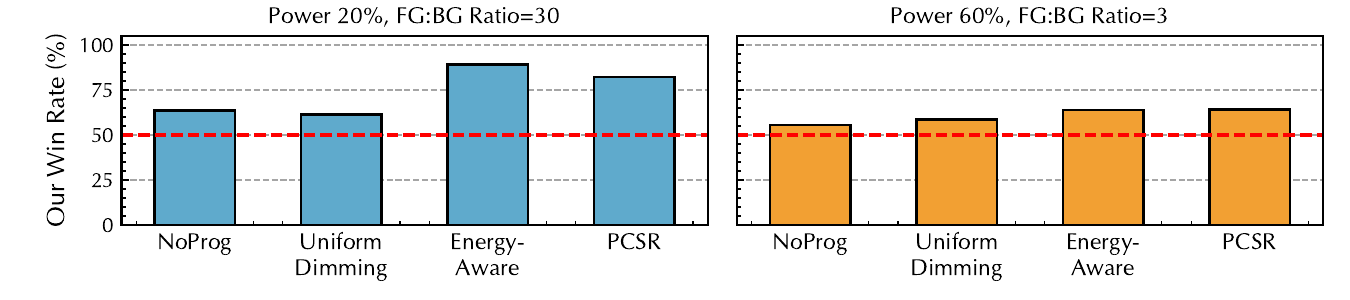}
    \caption{\textbf{User study results.} 2AFC results showing the average win rates of our method (across all participants and all trials) against 4 baselines. 
    50\% indicates a tie.
    Our method consistently outperforms all baselines ($p < 0.001$ in all cases; one-sided binomial test). 
    }

    \label{fig:user_study}
\end{figure*}

\paragraph{Distillation Details.}
For ground truth generation, we run iterative optimization on the training set at power targets from 90\% to 20\% with a step size of 10\%, across three FG:BG ratios ratios (3, 10, and 30), resulting in 4,800 training data.
We extend the HDRNet backbone channels by $2\times$ to improve fitting capability.
For training, we use a batch size of 8, learning rate of $10^{-4}$, and train for 60,000 iterations.

\paragraph{Baselines.}
We compare against three prior approaches to power-constrained display mapping, neither of which is specifically designed for AR displays:
\begin{itemize}
    \item \textbf{Uniform-Dimming} (\mode{UD}): it dims all pixel luminance by a global factor to meet the power budget.
    Chen et al.~\cite{chen2024pea} show that, under the same power consumption, \mode{UD} produces subjectively better results than other display mapping strategies that modulate pixel chromaticity.
    \item \textbf{Learned Per-Pixel Dimming Map}: we compare against Energy-Aware Images~\cite{le2023energy} (\mode{EAI}), a recent display mapping method that trains a convolutional neural network (CNN) to learn a per-pixel dimming map to meet the power target, similar to other prior works~\cite{nugroho2022r, le2023deep, itu2024display} and, more recently, \mode{ML-PEA}~\cite{chen2026mlpea}.
We choose \mode{EAI} as a comparison point over others for two reasons.
First, \mode{EAI} is much more light-weight than others, making is practical for AR usages (see \Sect{sec:exp:runtime}).
Second, \mode{EAI} takes an arbitrary target power ratios to train a single model, as opposed to training separate models for different power targets~\cite{chen2026mlpea}.
    
    \item \textbf{Learned Local Mapping}: we compare against \mode{PCSR}~\cite{yin2019power}, a local tile-based
    mapping method formulates tone mapping as a regularized sparse coding problem, jointly optimizing for power saving and perceptual quality.
    The mapping algorithms are optimized offline, not amenable for real-time processing.
\end{itemize}

\subsection{User Study}
\label{sec:exp:user}
We conduct an IRB-approved user study to validate that our method produces perceptually superior results compared to baselines under the same power budget.

\paragraph{Procedure.}
In addition to the three baselines described in~\Sect{sec:exp:setup}, we also compare against an iterative optimization variant without progressive training~(\Sect{sec:loss}), denoted as \mode{NoProg}, to show the importance of progressive training.
\mode{NoProg} provides an AR-specific controlled comparison: it uses the same OST-AR image formation, display model, TMO parameterization, and power constraint as our method, differing only in the progressive optimization strategy.

We conduct a Two-Alternative Forced Choice (2AFC) study against four baselines, similar to that used in a previous AR perception study~\cite{chapiro2024ar}.
In each trial, participants view three images displayed side by side: the original foreground image (reference) at the center, with tone-mapped results from our method and a baseline randomly placed on the left and right.
The backgrounds are displayed behind the foreground content for both our method and the baselines.
Participants are asked to freely view the images and select whether the left or right image is more similar to the reference at the center.

We use all 7 foregrounds and 7 backgrounds from the evaluation set, forming 49 foreground-background pairs (see Supplementary Material \ref{sec:supp:dataset} for details).
For each pair, we test two power budgets (20\% and 60\%).
To ensure that the foreground remains visible, we pair each power level with a suitable FG:BG luminance ratio: 30:1 for 20\% power and 3:1 for 60\% power, yielding $49 \times 2 = 98$ unique conditions.
With 4 baselines and left--right counterbalancing (i.e., swapping the placement of our method and the baseline), each participant completes $98 \times 4 \times 2 = 784$ comparisons, taking approximately one hour.
We recruit  10 participants (ages 18--35) with normal or corrected-to-normal vision.
By aggregating the responses across all participants, we obtain 1,960 pairwise comparisons against each baseline.

\begin{figure*}[!t]
    \centering
    \includegraphics[width=0.9\textwidth, alt={A foreground and background example followed by a grid comparing LowPowAR, NoProg, uniform dimming, EAI, and PCSR at two foreground-to-background luminance ratios and two power budgets.}]{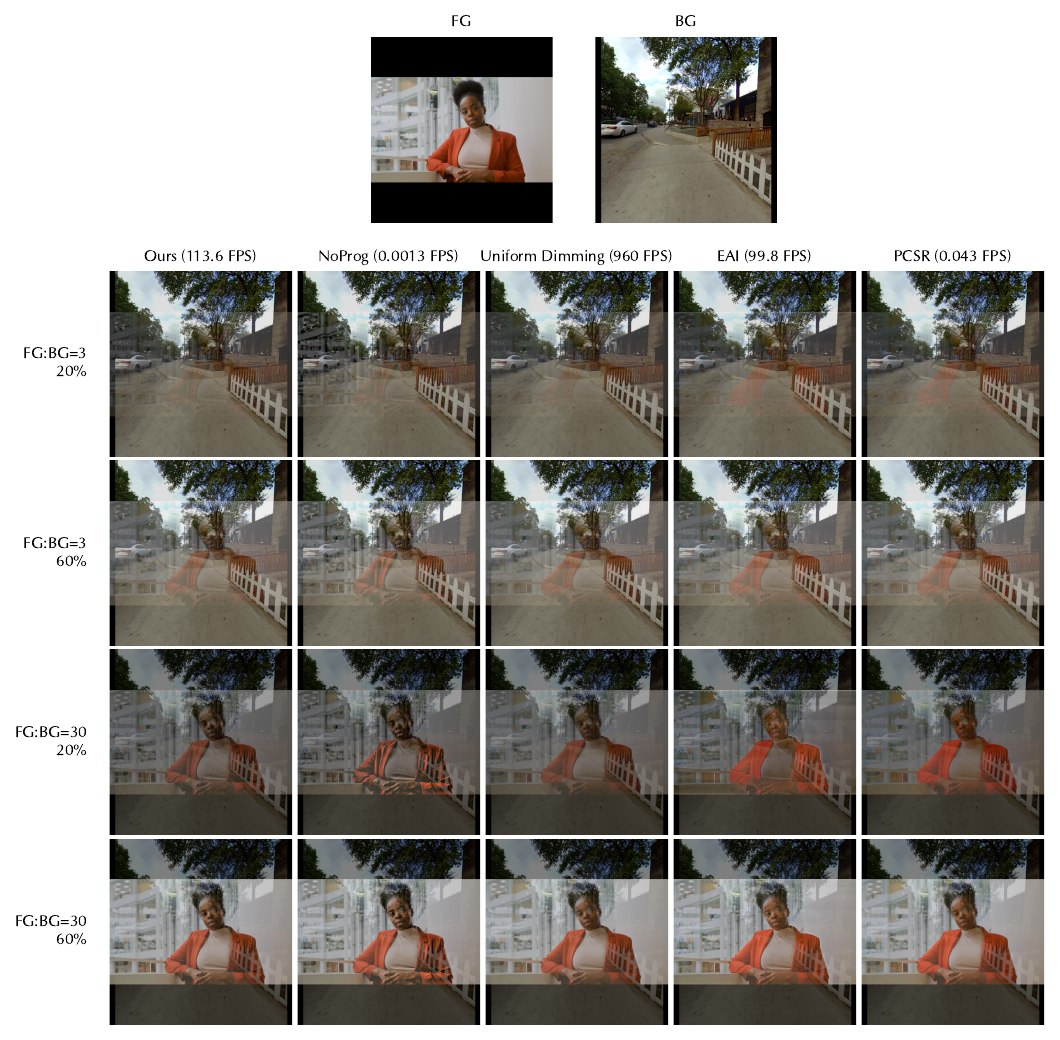}
    \caption{Visual comparison of our method against baselines across different power budgets and FG:BG ratios.
    Foreground peak luminance is set to 360 nits. More results are in Supplementary Material Figs.~\ref{fig:supp_blog_aria2}--\ref{fig:supp_phone_deadleaves}.
    The corresponding FPS values annotated in the figure were measured at $600 \times 600$ resolution on an NVIDIA Jetson Xavier AGX. }
    \label{fig:comparison}
\end{figure*}

\paragraph{Results.}
\Fig{fig:user_study} shows the percentage of times users prefer our method over each baseline under two viewing conditions.
Aggregating both viewing conditions, our method is selected in 76.4\% of comparisons against \mode{EAI} (95\% Wilson CI: 74.5\%--78.3\%; Cohen's $h=0.56$), 73.3\% against \mode{PCSR} (71.3\%--75.2\%; $h=0.49$), 60.0\% against \mode{UD} (57.8\%--62.1\%; $h=0.20$), and 59.6\% against \mode{NoProg} (57.4\%--61.7\%; $h=0.19$), with $n=1{,}960$ judgments per comparison.
All results are statistically significant ($p < 0.001$, one-sided binomial test).
Our method is preferred under both viewing conditions.
The advantage is especially pronounced at low display power (20\% power, FG:BG\,=\,30:1), where win rates reach 89.1\% (vs.\ \mode{EAI}), 82.3\% (vs.\ \mode{PCSR}), 63.7\% (vs.\ \mode{NoProg}), and 61.4\% (vs.\ \mode{UD}).
At higher power (60\%, FG:BG\,=\,3:1), our method still outperforms all baselines, with win rates of 64.3\%, 63.8\%, 58.6\%, and 55.5\% against \mode{PCSR}, \mode{EAI}, \mode{UD}, and \mode{NoProg}, respectively
(all $p < 0.001$).

\paragraph{Visual Comparison.}
To allow a more direct assessment, we visualize our method alongside all baselines in \Fig{fig:comparison} using one FG/BG combination, showing results across two FG:BG luminance ratios and two display power budgets.
See Supplementary Material~\ref{sec:supp:comparisons} for more visualizations.
These visual observations are consistent with the user study results presented in \Sect{sec:exp:user}.
\mode{EAI} and \mode{PCSR} produce noticeable non-uniform artifacts, particularly at low display power (20\%).
\mode{UD} appears smooth and natural overall but suffers from insufficient contrast.
\mode{NoProg} tends to exaggerate contrast (e.g., the jacket and human hands in the case with FG:BG\,=\,30:1 and 20\%  power).
\mode{Ours} strikes a good balance between enhancing contrast and maintaining natural appearance without exaggerating contrast.

\paragraph{Evaluation Using Existing Perceptual Metrics.}
While no perceptual metrics exist specifically for low-power, additive AR displays, we can still leverage our user study data as ground truth to carry out a first-order analysis of existing image quality metrics.
We evaluate three popular metrics (ColorVideoVDP~\cite{mantiuk2024colorvideovdp}, MILO~\cite{cogalan2025milo}, and LPIPS~\cite{zhang2018unreasonable}) across different FG:BG ratios and power targets, using uniform backgrounds for additive compositing as suggested by Chapiro et al.~\cite{chapiro2024ar}.

The detailed results, including an additional PQ-encoded~\cite{smpte2014st2084} evaluation, are shown in Supplementary Material \ref{sec:supp:quant}.
Briefly, ColorVideoVDP best agrees with our user study results by consistently predicting that our method achieves the best quality over baselines.
In contrast, MILO and LPIPS both favor \mode{EAI}, contradicting the user study data; this mismatch persists after PQ encoding.
ColorVideoVDP predicts that all baselines are within 0.2 Just Objectionable Difference (JOD) from our method. 
This contradicts our user study results, where our win rate against \mode{EAI} and \mode{PCSR} exceeds 75\%, indicating a true perceptual difference of over 1 JOD.
These findings highlight the need for perceptual metrics designed specifically for AR displays~\cite{zhang2021perceived, zhang2018color}, especially under power-constrained conditions.

\subsection{Distillation for Real-time Performance}
\label{sec:exp:runtime}

We distill the iterative optimization into a lightweight feedforward network for real-time inference (\Sect{sec:realtime}).
We first validate that our distilled network successfully approximates the iterative optimization results.
The PSNR converges to over 47\,dB (see Supplementary Material \ref{sec:supp:training} for training curve), indicating that the distilled network produces results close to iterative optimization.

We measure the speed of the distilled pipeline on an NVIDIA Jetson Xavier AGX~\cite{jetson_agx_xavier}, an embedded GPU representative of XR-capable mobile hardware.\footnote{The peak GPU throughput is approximately 1.4\,TFLOPS in FP32.
For comparison, the Snapdragon XR2 Gen~2 SoC used in Meta Quest~3 integrates an Adreno~740 GPU, whose peak FP32 throughput is estimated to be $\sim$2\,TFLOPS~\cite{adreno740_wikipedia}.}
At $600 \times 600$ resolution that matches current AR display~\cite{metarayban2025}, our method achieves 8.8\,ms end-to-end latency (113.6\,FPS).
Supplementary Material~\ref{sec:supp:runtime} reports the latency breakdown.
For reference, we also measure the throughput of the baselines on the same device: \mode{EAI} runs at 99.8\,FPS, \mode{PCSR} at only 0.043\,FPS due to its iterative optimization process, and \mode{ML-PEA}  at 3.3\,FPS due to its heavy U-Net architecture.

The latency of our method is even lower when the tone mapping is offloaded to the cloud/remote server. 
On a RTX 5090 GPU, the latency is reduced to only 0.7 ms.

\begin{figure*}[t]
    \centering
    \begin{subfigure}[b]{0.56\textwidth}
        \centering
        \raisebox{1mm}{\includegraphics[width=\textwidth, alt={Two visual comparisons showing that progressive optimization avoids exaggerated contrast produced by direct optimization.}]{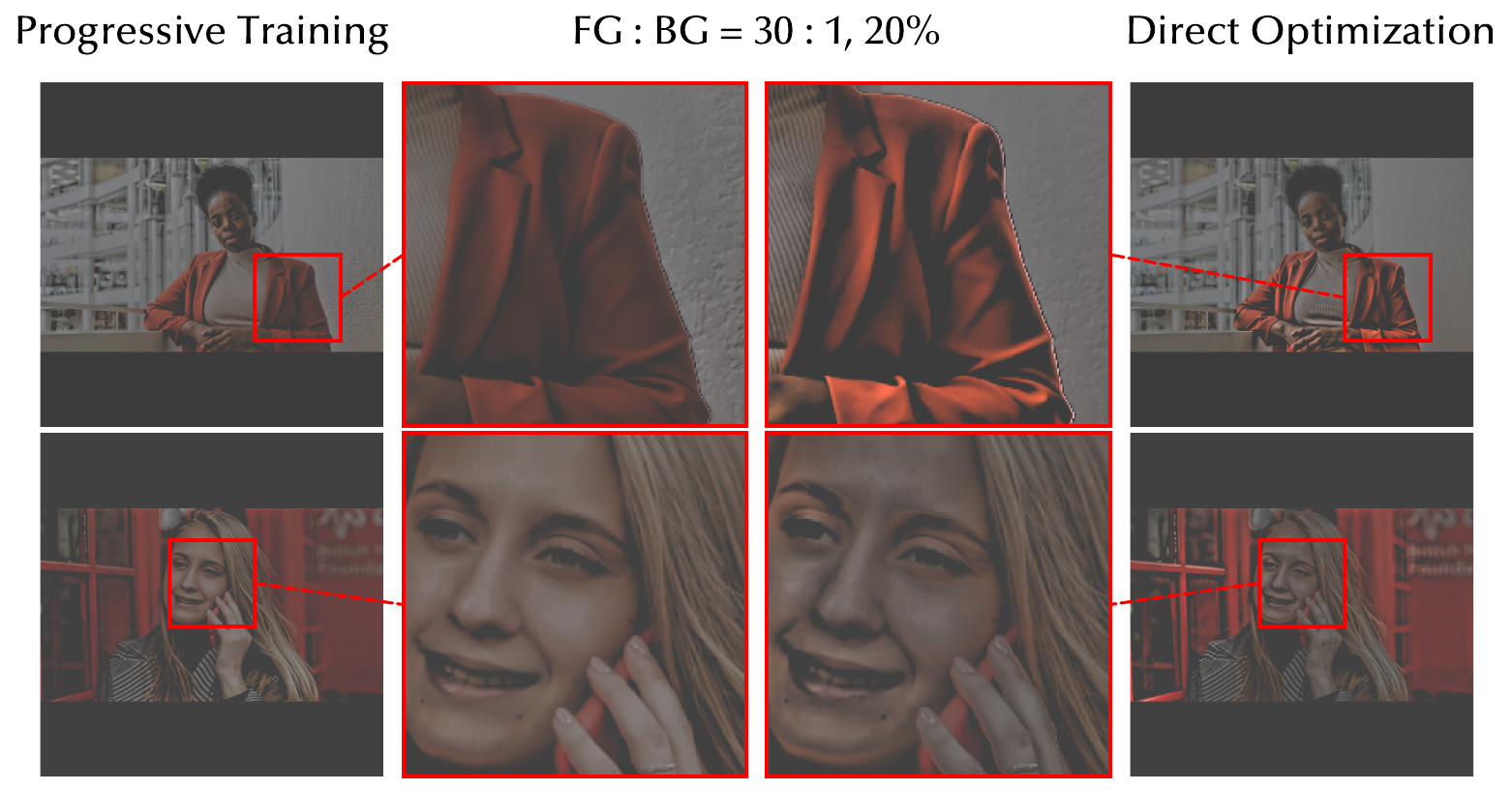}}
        \caption{Progressive optimization vs.\ direct optimization.}
        \label{fig:ablation_progressive}
    \end{subfigure}%
    \hfill
    \begin{subfigure}[b]{0.4\textwidth}
        \centering
        \raisebox{1.2mm}{\includegraphics[width=\textwidth, alt={A visual comparison in which direct network training contains artifacts and the distilled model resembles the iterative result.}]{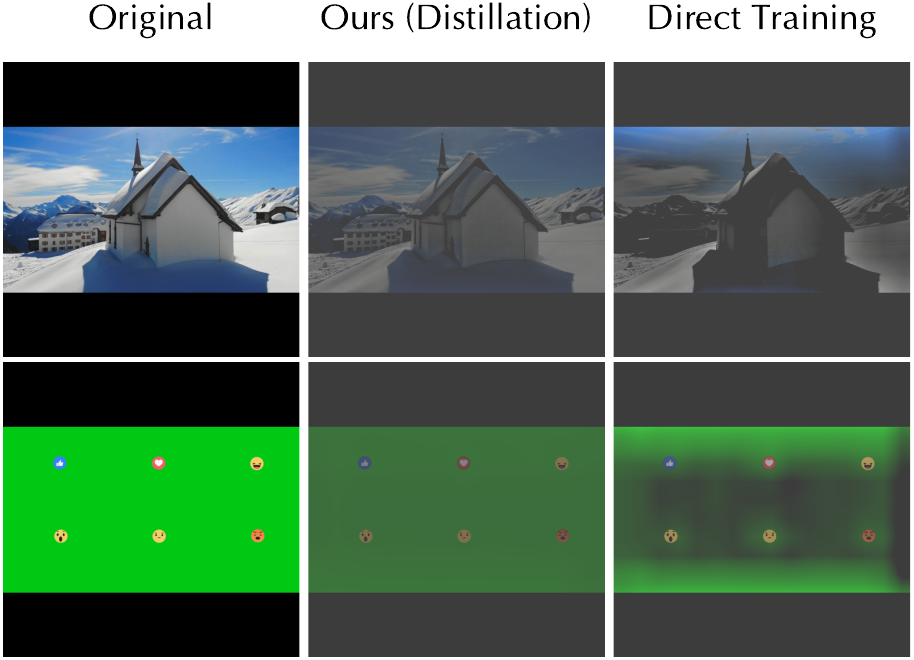}}
        \caption{Direct training vs.\ distillation.}
        \label{fig:ablation_distill}
    \end{subfigure}%
    \\[2mm]
    \begin{subfigure}[b]{0.56\textwidth}
        \centering
        \includegraphics[width=0.97\textwidth, alt={Results optimized with an average uniform background and with a precise spatially varying background under several luminance ratios and power budgets.}]{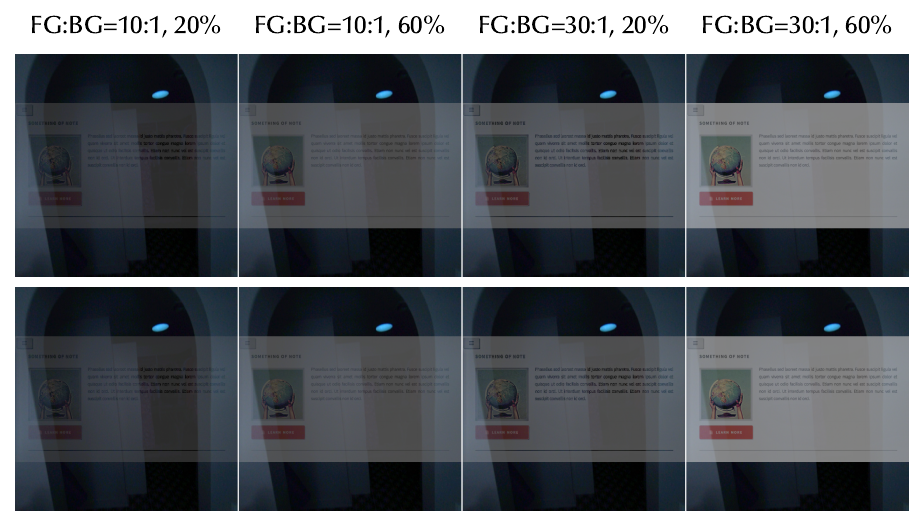}
        \caption{Uniform vs.\ precise background.}
        \label{fig:ablation_bg}
    \end{subfigure}%
    \hfill
    \begin{subfigure}[b]{0.4\textwidth}
        \centering
        \raisebox{2mm}{\includegraphics[width=\textwidth, alt={A visual comparison showing better detail preservation with bilateral base-detail decomposition than with direct piecewise tone mapping.}]{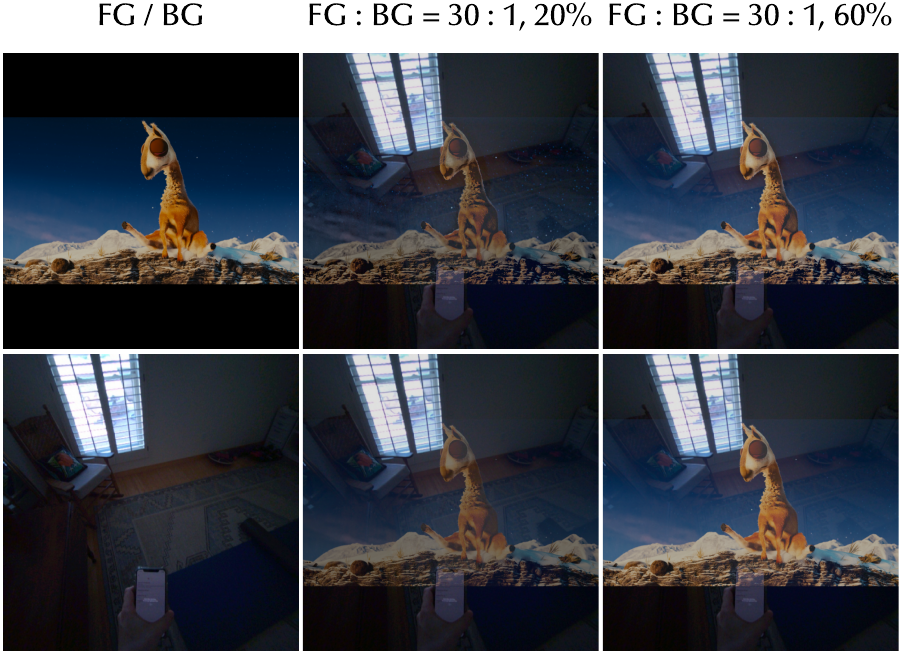}}
        \caption{Bilateral decomposition.}
        \label{fig:ablation_bd}
    \end{subfigure}%
    \caption{\textbf{Ablation studies.}
    \textbf{(a)}: Progressive optimization versus direct optimization to the target power. Direct optimization overly exaggerates the contrast, producing artifacts (more in Supplementary Material Fig.~\ref{fig:supp_progressive}).
    \textbf{(b)}: Directly training the inference network with ColorVideoVDP and power loss produces visible artifacts that our distillation approach avoids (FG:BG\,=\,30:1, 20\% power; more in Supplementary Material Fig.~\ref{fig:supp_distillation}).
    \textbf{(c)}: Results optimized using uniform background (top) versus precise background (bottom) across different FG:BG ratios and power budgets (more in Supplementary Material Fig.~\ref{fig:supp_uniform_precise_2}).
    \textbf{(d)}: Our method with bilateral decomposition (top) versus direct piecewise tone mapping (bottom). Decomposition better preserves high-frequency details, particularly at lower power budgets (20\%; more in Supplementary Material Fig.~\ref{fig:supp_decomp_1}).
    All use foreground peak luminance of 360 nits.}
    \label{fig:ablation_prog_bg}
\end{figure*}

\subsection{Extension to Dynamic Content and Background}
\label{sec:exp:video}

While our user study was done on static images over static background (like recent studies in AR perception~\cite{chapiro2024ar, kim2025supra}), our algorithm operates on a frame by frame basis and can, thus, be readily extended to scenarios where the foreground and/or background are dynamic.
We implement this strategy and show the videos in the Supplementary Material (see \texttt{video\_tmo\_examples.zip}), comparing our method with \mode{UD}.

We set the power target to 20\%.
We use videos from GUI-World~\cite{chen2025guiworld} as the foreground and egocentric videos from the Aria dataset~\cite{pan2023aria} as the background.
We globally scale the background video so that the overall FG:BG ratio is 30:1, and resample all videos to 30 FPS.
Since the background changes over time, we use a moving window of the past 30 frames (one second) to estimate the mean background RGB value (i.e., $\bar{I}_{bf}$ in \Sect{sec:realtime}).
Our results give better subjective quality than \mode{UD}.
In addition, no temporal flickering is observed in our results, confirming that our per-frame method can be directly applied to video.

An interesting direction in the future is to exploit the temporal dimension in dynamic content, which could provide additional opportunities to reduce display power by, for instance, accounting for the temporal sensitivity of the human visual system~\cite{ashraf2024castlecsf, cai2024elatcsf} and the time course of adaptation~\cite{chen2025modeling, pattanaik2000time, pattanaik1998multiscale, ferwerda1996model, yee2001spatiotemporal, 10.2312:sr.20251183}.

\subsection{Ablation Studies}
\label{sec:exp:ablation}

\subsubsection{Progressive Optimization Strategy}
\label{sec:exp:ablation:progressive}

\begin{figure*}[t]
    \centering
    \begin{subfigure}[b]{0.64\textwidth}
        \centering
        \includegraphics[width=\textwidth, alt={Visual comparison of the proposed tone-mapping parameterization with gamma, Laplacian-pyramid, and Reinhard alternatives at two power budgets.}]{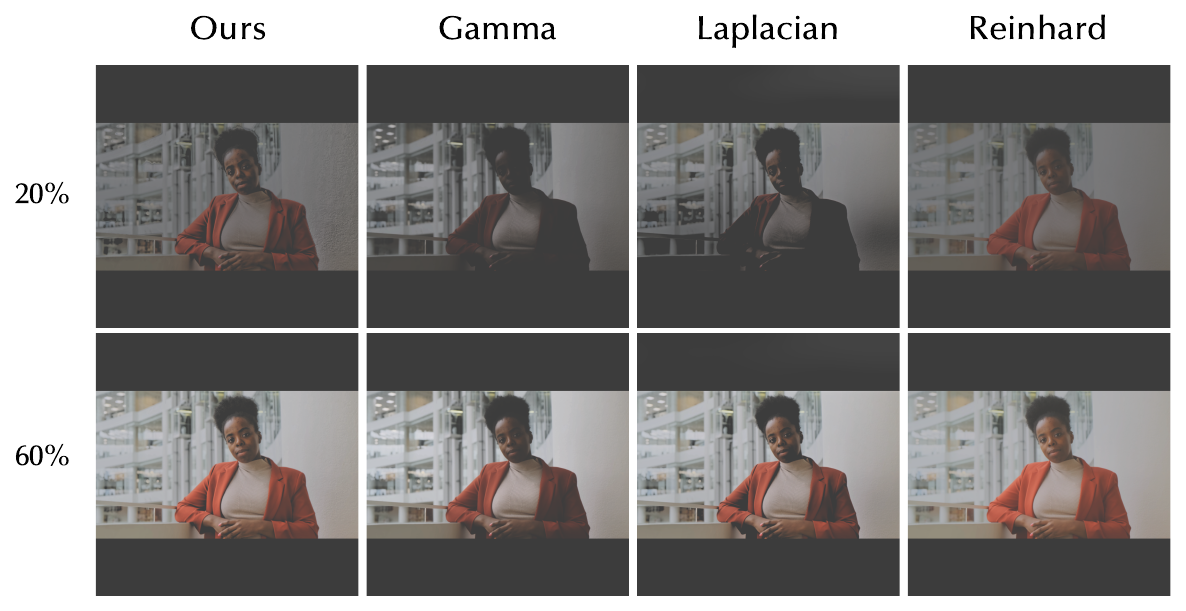}
        \caption{TMO parameterizations with learnable parameters.}
        \label{fig:ablation_tmo}
    \end{subfigure}%
    \hfill
    \begin{subfigure}[b]{0.30\textwidth}
        \centering
        \raisebox{-0.7mm}{\includegraphics[width=\textwidth, alt={Two examples comparing LowPowAR with simple base-layer scaling, which produces unnatural local contrast.}]{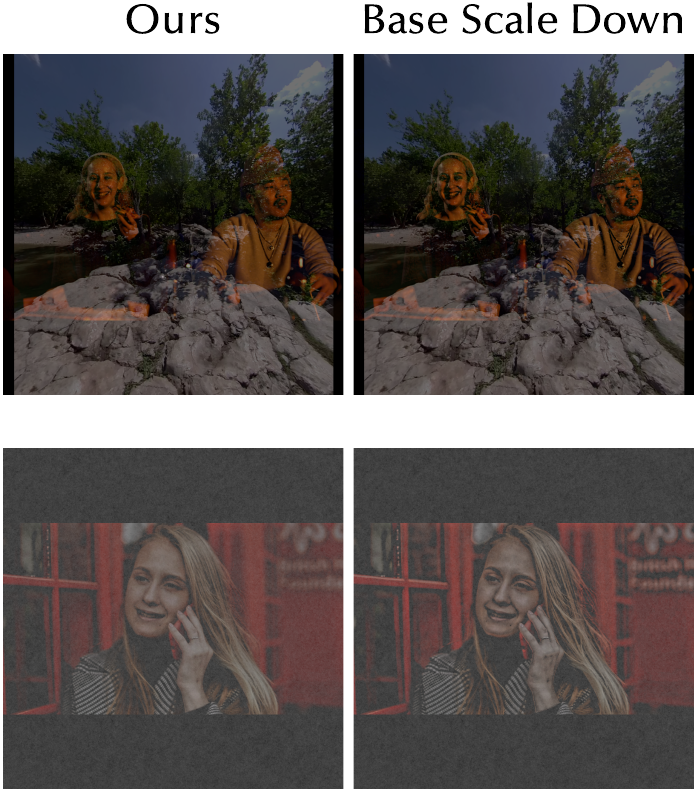}}
        \caption{Ours vs.\ simple base-layer scaling.}
        \label{fig:ablation_base_scale_down}
    \end{subfigure}%
    \caption{\textbf{Ablation studies for TMO choice and base layer scaling.}
    \textbf{Left}: Comparison of different TMO parameterizations under two power budgets (top: 20\%; bottom: 60\%) and 30:1 FG:BG ratio. Our approach preserves details better than alternative TMOs (more in Supplementary Material Fig.~\ref{fig:supp_tmo}).
    \textbf{Right}: Base layer scaling (right) amplifies local contrast, producing unnatural results. Our method (left) preserves natural appearance. FG:BG\,=\,30:1, 20\% power.
    All use foreground peak luminance of 360 nits.}
    \label{fig:ablation_tmo_base}
\end{figure*}

Our user study~(\Sect{sec:exp:user} and \Fig{fig:user_study}) confirms that our progressive optimization (\Sect{sec:loss:opt}) outperforms the variant without progressive training.
\Fig{fig:ablation_progressive} gives another visual example that compares the two variants at a 20\% power target and FG:BG ratio of 30:1: our method with progressive optimization (left) versus direct optimization using the original image at 100\% power as reference (right).
The latter exaggerates the contrast and produces visible artifacts.
Our progressive strategy reduces these artifacts by choosing appropriate references at each optimization step.
We show more examples in Supplementary Material Fig.~\ref{fig:supp_progressive}.

In Supplementary Material Section~\ref{sec:supp_perceptual}, we show that this issue is not specific to ColorVideoVDP~\cite{mantiuk2024colorvideovdp}, but is a general problem when using existing perceptual metrics to compare images with large appearance discrepancies.
For instance, the recent MILO metric~\cite{cogalan2025milo} suffers from the same issue, producing different but equally problematic artifacts.

\subsubsection{Direct Training vs. Distillation}
\label{sec:exp:ablation:dt}

As discussed in \Sect{sec:realtime}, we adopt a distillation approach rather than directly training the network with perceptual loss and power constraints.
To validate this design choice, we compare two training strategies:
(1) \textbf{Direct Training}: training the network directly using $\mathcal{L}_{\text{perceptual}}$ and $\mathcal{L}_{\text{power}}$;
(2) \textbf{Ours (Distillation)}: training with ground truth generated by our offline iterative optimization.
\Fig{fig:ablation_distill} shows the comparison with two examples.
See Supplementary Material Fig.~\ref{fig:supp_distillation} for more examples.
Direct training tends to be unstable, converging to suboptimal solutions with visible artifacts, whereas our distillation approach yields significantly better outputs.

\subsubsection{Background Representation}
\label{sec:exp:ablation_bg}

Instead of the precise background image, we use its average luminance to derive the TMOs (see \Eqn{eq:ar_composite_2} and \Sect{sec:problem}). 
\Fig{fig:ablation_bg} compares results optimized with uniform backgrounds (top) versus precise backgrounds (bottom) using our iterative optimization
(more examples in Supplementary Material Fig.~\ref{fig:supp_uniform_precise_2}).
The two approaches produce similar results.
Our results align with prior work~\cite{chapiro2024ar} showing that averaged background luminance best predicts user preference in additive displays.

\subsubsection{Bilateral Decomposition}
\label{sec:exp:ablation:bd}

Our method applies separate tone mapping to the base and detail layers from bilateral decomposition.
We compare against a variant that learns the piecewise-linear TMOs (\Sect{sec:tonemapper:tmo}) directly to the original image luminance without base/detail separation.
\Fig{fig:ablation_bd} shows an example that compares our method (top) versus one without bilateral decomposition (bottom).
Our method better preserves high-frequency details because it explicitly manipulates high-frequency details.
The results are consistent with other examples, which we show in Supplementary Material Fig.~\ref{fig:supp_decomp_1}.

\subsubsection{TMO Choice}
\label{sec:exp:ablation:tmo}

We compare our bilateral decomposition-based TMO against alternative learnable tone mapping formulations.
We evaluate four approaches:
(1) \textbf{Ours}: described in \Sect{sec:tonemapper};
(2) \textbf{Gamma}: per-anchor learnable gamma parameter;
(3) \textbf{Laplacian Pyramid}: Laplacian pyramid decomposition~\cite{burt1987laplacian} following Mertens et al.~\cite{mertens2007exposure} used in HDR+~\cite{hasinoff2016burst}, with per-anchor learnable weights for each frequency band;
(4) \textbf{Reinhard}: photographic tone mapping~\cite{reinhard2002photographic} with per-anchor learnable local adaptation parameter $\alpha$.
All variants are optimized using our progressive iterative optimization (\Sect{sec:loss:opt}).

\Fig{fig:ablation_tmo} shows the comparison with an example with a FG:BG ratio of 30:1 and two power budgets at 20\% and 60\%.
See Supplementary Material Fig.~\ref{fig:supp_tmo} for more examples.
The Gamma and Laplacian Pyramid methods tend to exaggerate contrast while Reinhard results are dull with weak contrasts, especially at low power.

\subsubsection{Inference Pipeline Design}
\label{sec:ablation:pipeline}
Our real-time inference pipeline (\Sect{sec:realtime}) imitates the offline, iterative optimization: it applies separate TMOs to base and detail layers and operates only on luminance.
We validate this design choice by comparing against two variants:
(1) \emph{without bilateral decomposition}: where the CNN predicts the luminance scaling map directly for the original image without base/detail separation.
(2) \emph{without luminance-only constraint}: where the CNN predicts full RGB affine transformations (as in HDRNet~\cite{gharbi2017deep}).

Using the iterative optimization results as the ground truth, we measure PSNR on the evaluation set.
Our full pipeline achieves 47.62\,dB.
Removing bilateral decomposition drops PSNR to 45.99\,dB ($-$1.63\,dB), as the network struggles to fit high-frequency details without explicit decomposition.
Removing the luminance-only constraint yields 47.49\,dB ($-$0.13\,dB), showing that learning a generic affine transformation is unnecessary.

\subsubsection{Simple Base Layer Scaling Down}
\label{sec:exp:ablation:base_scale_down}

We also compare against a simple, non-learned alternative, where we simply scale down the base layer according to the target power budget, then add back the original detail layer.
\Fig{fig:ablation_base_scale_down} shows examples at FG:BG\,=\,30:1 and 20\%  power.
Since the base layer (captures local mean) is reduced while the detail layer (captures local variation) remains unchanged, the local contrast (variation divided by mean) is amplified, which could lead to unnatural appearance, e.g., the faces in both examples.
In contrast, our method jointly adjusts base and detail layers, achieving a natural appearance.

\section{Discussion and Limitations}
\label{sec:exp:limitation}

\paragraph{Controlled beam-splitter validation versus real AR deployment.}
Our beam-splitter setup follows prior controlled optical see-through AR (OST-AR) perception studies~\cite{murdoch2020brightness,zhang2021perceived,herbeck2024transparency,chapiro2024ar,kim2025supra}.
It provides a controlled testbed in which foreground and background luminance---two variables critical to tone mapping---can be independently and precisely controlled, which is difficult to achieve with real AR glasses in natural environments.
This prototype does not reproduce all factors encountered in commercial glasses or real-world deployment, such as waveguide artifacts, eyebox variation, device-calibration drift, and natural illumination.
AR/MR device measurements are also sensitive to environmental conditions~\cite{guo2022measurement}.
Nevertheless, these factors can be incorporated into our framework through richer forward models.

For example, the image-space non-uniformity map used by AR-DAVID provides a phenomenological model of waveguide artifacts~\cite{chapiro2024ar}, while first-principles approaches, such as the method described by Kress et al.~\cite{kress2020optical}, can model waveguide optics more accurately.
An eye model can additionally account for eyebox variation by mapping the optical output to the retinal image.
These extensions retain the constrained-optimization structure of our method while enabling power reduction under more realistic deployment conditions.

\paragraph{Computational Overhead}
Our method incurs modest computational overhead (113.6 FPS). 
When on-glasses compute resources are limited, inference can be offloaded to a paired smartphone or edge device, further reducing the computational burden on the glasses.

\paragraph{Small Text Artifacts.}
We observe that when the foreground contains small dark text on a white canvas, our tone mapping can introduce noise that reduces text legibility; see Supplementary Material Fig.~\ref{fig:supp_limitation} for examples.
We attribute this artifact to the high spatial frequency and high contrasts of the sharp edges in texts.
Text processing (e.g., compression) is known to require different algorithms than natural images~\cite{uchigasaki2023deep, della2025end}.
In general, an interesting future work is to consider the semantic information in the content, e.g., human faces vs. natural landscape vs. texts when optimizing the TMOs.

\section{Conclusion}
\label{sec:conclusion}

We demonstrate an effective approach to reduce AR display power through the lens of power-constrained tone mapping.
Our contributions are 1) an optimization-friendly parameterization of the TMOs that can flexibly trade quality for power, 2) a progressive optimization strategy that side-steps adversarial stimuli, and 3) a distillation method to enable real-time deployment.

\section*{Supplemental Materials}
The supplement provides additional details on the display model, failure cases of existing metrics, extended results, ablation studies, further comparisons, and examples on video content.

\section*{Figure Credits and Copyrights}
All figure compositions are by the authors; source imagery is from the AR-DAVID~\cite{chapiro2024ar}, XR-DAVID~\cite{chapiro2024xr}, Aria~\cite{pan2023aria}, and PEA-PODs~\cite{chen2024pea} datasets.

\acknowledgments{%
The work is partially supported by NSF Award \#2225860 and a Meta research grant.
}

\clearpage

\bibliographystyle{abbrv-doi-hyperref}
\bibliography{references}

\clearpage
\setcounter{section}{0}
\renewcommand{\thesection}{\Alph{section}}
\makeatletter
\twocolumn[{
    \centering

    {\huge\sffamily Supplementary Material for ``LowPowAR: Power-Constrained Tone Mapping for Augmented Reality''\par}
    \vspace{30pt}

    {\large\sffamily
      \begin{tabular}[t]{@{}c@{}}
        \vgtc@author
      \end{tabular}\par}
    \vspace{30pt}

    \includegraphics[width=\textwidth, alt={Examples in which direct optimization exploits ColorVideoVDP or MILO under a large reference-to-test appearance gap, producing luminance non-uniformity or high-frequency noise despite a lower metric loss.}]{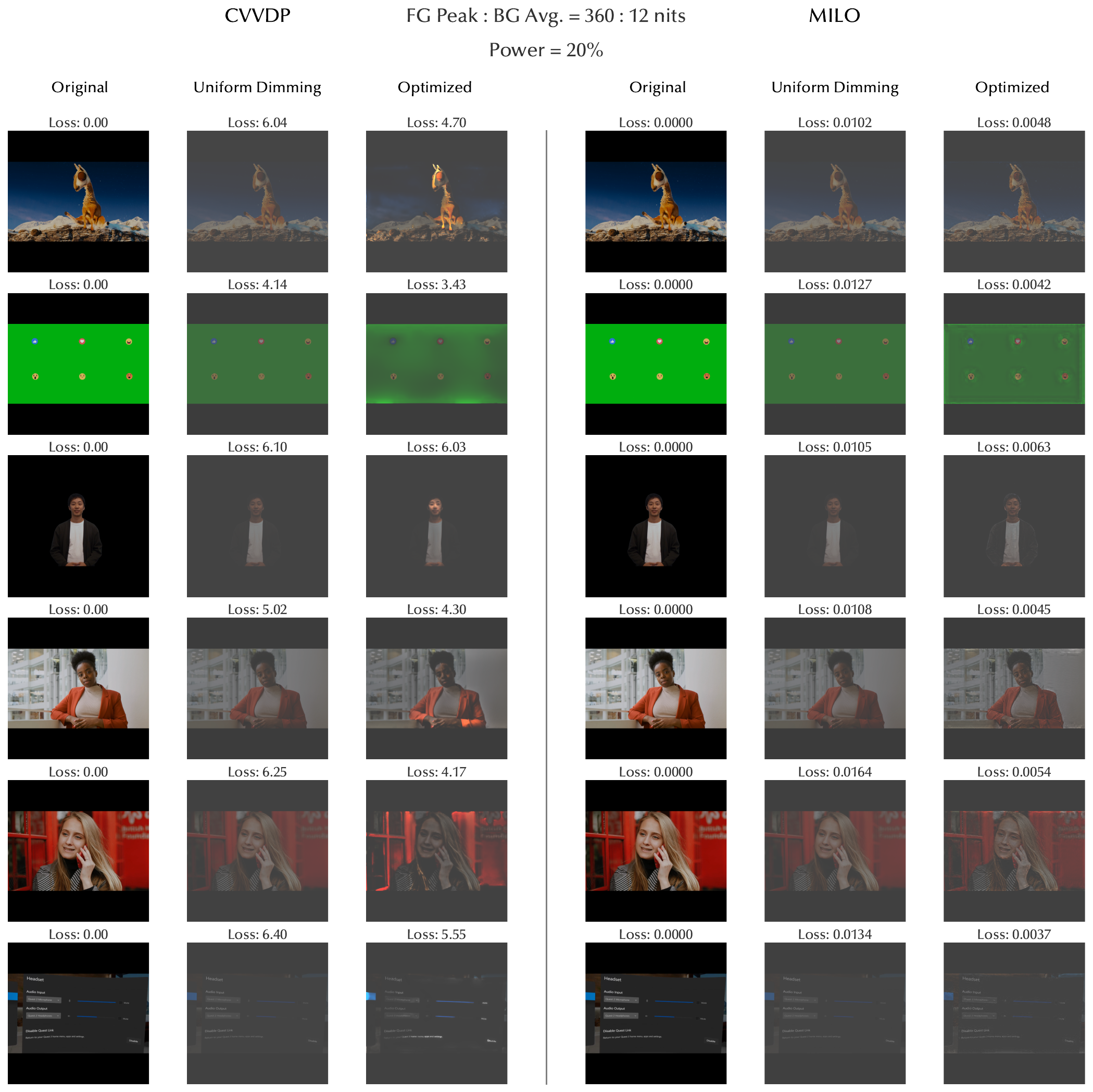}
    \refstepcounter{figure}
    \vspace{2pt}
    {\captionfonts\raggedright\noindent Figure~\thefigure.
    \textbf{Failure modes of perceptual metrics under large appearance gaps.}
    ColorVideoVDP optimization (left) yields luminance non-uniformity, while MILO (right) yields high-frequency noise, despite lower metric loss than uniform dimming.\par}
    \label{fig:supp_perceptual}
    \vspace{1em}
}]
\makeatother

\section{Display Model}
\label{sec:supp:display_model}
We use the standard gamma-offset-gain (GOG) model~\cite{mantiuk2024colorvideovdp, mantiuk2008display} to convert pixel values to physical luminance.
Given an sRGB pixel value $V \in [0, 1]$, the display model $\mathcal{D}$ computes the emitted luminance $L$ as follows:

\paragraph{Step 1: Gamma Decoding.}
We first convert sRGB values to linear light using the sRGB transfer function:
\begin{equation}
    V_{\text{linear}} =
    \begin{cases}
        V / 12.92 & \text{if } V \leq 0.04045 \\
        \left( \frac{V + 0.055}{1.055} \right)^{2.4} & \text{otherwise}
    \end{cases}
\end{equation}

\paragraph{Step 2: Luminance Mapping.}
The linear value is then mapped to physical luminance using the display's black level $L_{\min}$ and peak luminance $L_{\max}$:
\begin{equation}
    L = L_{\min} + (L_{\max} - L_{\min}) \cdot V_{\text{linear}}
\end{equation}

Our foreground monitor has a peak luminance of 1,200 nits and a contrast ratio of 2000:1.
After reflection through the 70:30 beam splitter, the effective peak luminance is $L_{\max} = 360$ nits.
The black level is derived from the contrast ratio: $L_{\min} = L_{\max} / 2000 = 0.18$ nits.
Similar modeling is applied to our background monitor.

\section{Perceptual Loss Under Large Appearance Gaps}
\label{sec:supp_perceptual}

We demonstrate the failure modes of prior perceptual losses when optimizing images with large appearance differences between the reference and test images.
As discussed in the main paper (\Sect{main-sec:problem}), this situation arises in our setting due to: (1) aggressive display power reduction and (2) additive AR background.

\subsection{Experimental Setup}

We use two advanced perceptual losses: ColorVideoVDP~\cite{mantiuk2024colorvideovdp} and MILO~\cite{cogalan2025milo}.
Both are used to minimize the perceptual difference between the power-reduced image (composited with the background) and the original image.

To demonstrate the failure modes, we use a target power of 20\% and an FG:BG ratio of 30:1, which is a challenging setting with significant appearance change.
To provide maximum optimization freedom and thus exposing the most artifacts when perceptual metrics fail,
we parameterize the tone mapping as per-pixel scaling, reduce the power contraint lambda $\lambda_p$ (1000 for ColorVideoVDP and 10 for MILO), and disable all regularization terms (smoothness loss and progressive training).

\subsection{Results}

\Fig{fig:supp_perceptual} shows representative examples comparing the original image, uniform dimming, and per-pixel scaling optimization using ColorVideoVDP (left) and MILO (right).
When optimizing directly against the original image without strong regularization, both perceptual losses produce severe artifacts, albeit in different forms: ColorVideoVDP leads to luminance non-uniformity (\Fig{fig:supp_perceptual}, left), while MILO introduces high-frequency noise (\Fig{fig:supp_perceptual}, right).
This occurs because under the combined influence of the power constraint and the additive background, these perceptual metrics provide unreliable gradients that can be exploited by gradient descent.
Each metric is ``hacked'' in a different way, leading to adversarial-like solutions that minimize the loss but result in visually unacceptable outputs.

\section{Dataset Overview}
\label{sec:supp:dataset}

\begin{figure*}[t]
    \centering
    \includegraphics[width=0.8\textwidth, alt={A grid of 17 foreground images in the evaluation set.}]{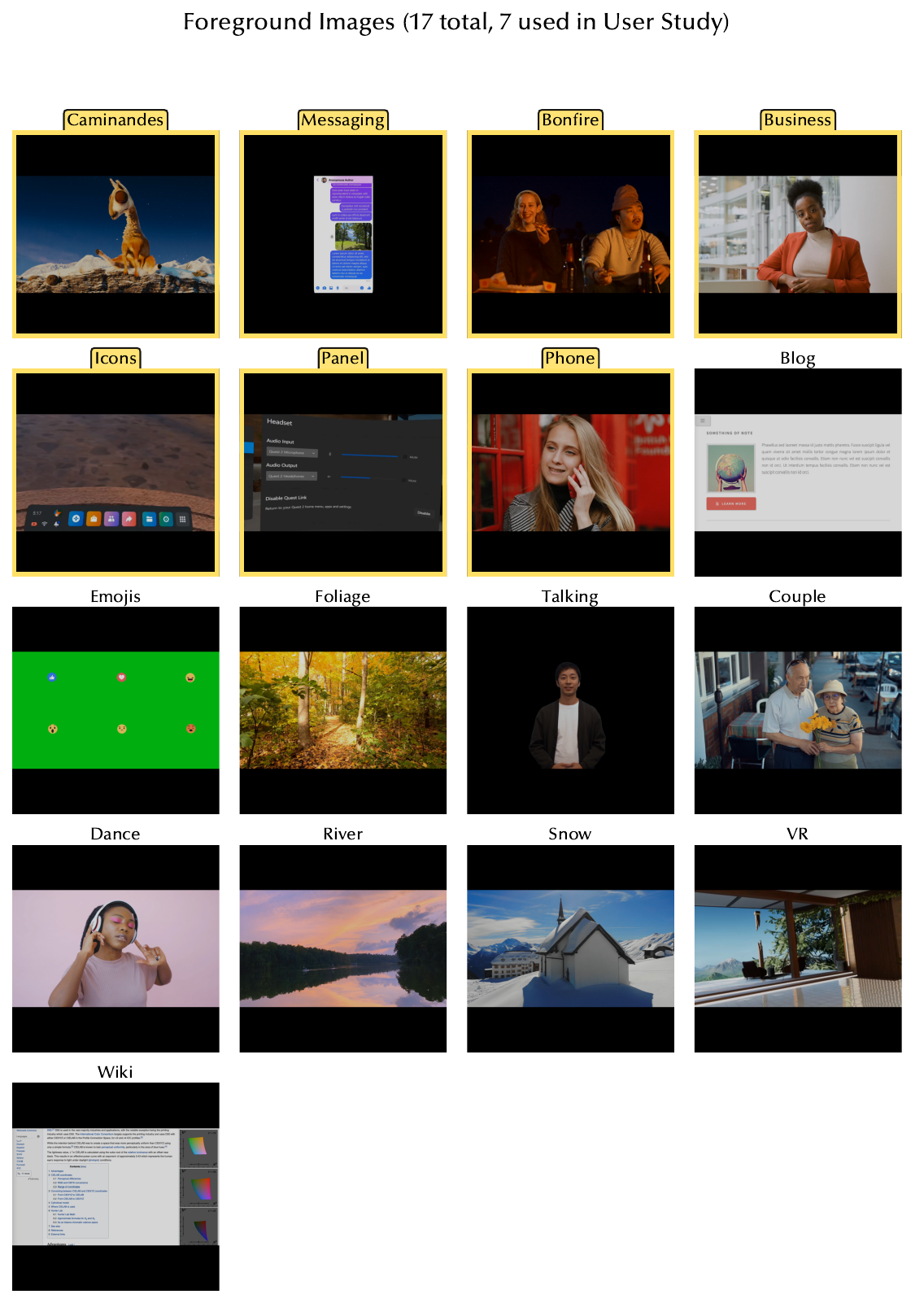}
    \caption{\textbf{Foreground images in the evaluation set.}
    17 foreground images covering diverse AR content: video calls, messaging apps, web pages, emojis, natural scenes, and UI elements.
    Yellow-highlighted images are used in the user study.
    }
    \label{fig:supp_foreground}
\end{figure*}

\begin{figure*}[t]
    \centering
    \includegraphics[width=0.75\textwidth, alt={A grid of 11 evaluation backgrounds.}]{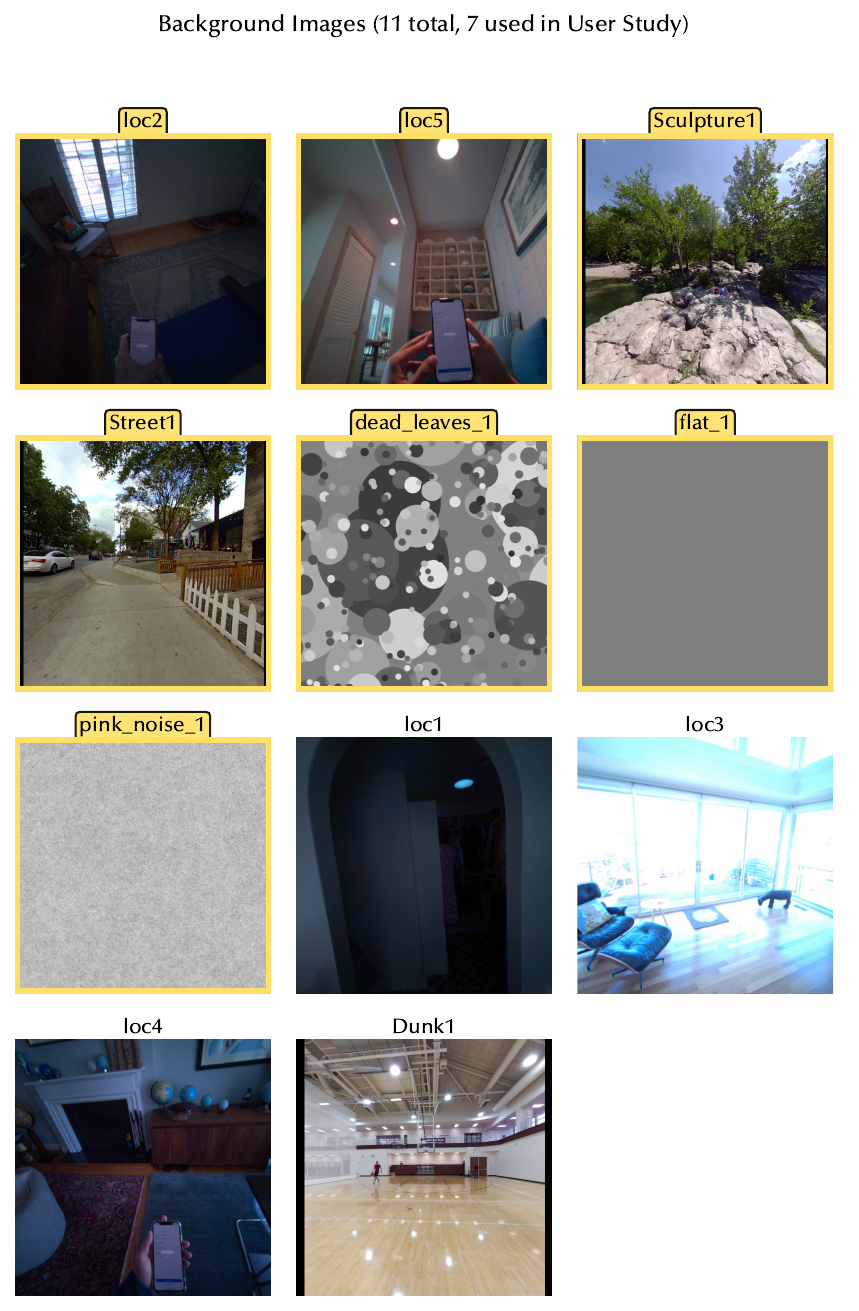}
    \caption{\textbf{Background images in the evaluation set.}
    11 background images: 5 egocentric captures from Aria (loc1--loc5), 3 outdoor scenes from PEA-PODs (Sculpture1, Street1, Dunk1), and 3 synthetic patterns (dead\_leaves, flat, pink\_noise).
    Yellow-highlighted images are used in the user study.
    }
    \label{fig:supp_background}
\end{figure*}

\Fig{fig:supp_foreground} and \Fig{fig:supp_background} show all foreground and background images used in our evaluation.
The evaluation set contains 17 foreground images and 11 background images.
Yellow-highlighted images (7 foregrounds and 7 backgrounds) are used in the user study.

\begin{table}[t]
    \centering
    \caption{\textbf{Runtime breakdown} on Jetson Xavier AGX.}
    \label{tab:runtime}
    \resizebox{1\columnwidth}{!}{%
    \begin{tabular}{ccccc}
    \toprule
    Total & Decompos. & CNN backbone & Slicing & Others  \\
    \midrule
    8.8 ms & 36.8\% & 27.1\% & 25.4\% & 10.7\%  \\
    \bottomrule
    \end{tabular}}
\end{table}
\begin{figure}[t]
    \centering
    \includegraphics[width=\linewidth, alt={Training curves showing validation loss decreasing and PSNR increasing toward approximately 47.4 decibels over 60 thousand distillation iterations.}]{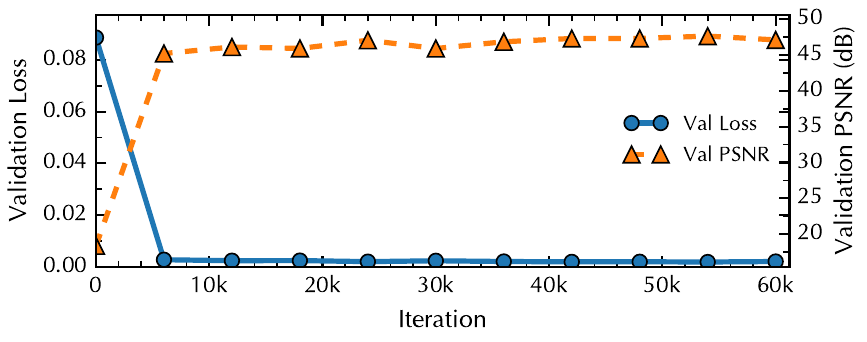}
    \caption{\textbf{Distillation training convergence.}
    Validation loss (blue, left axis) and PSNR (orange, right axis) over 60k training iterations.
    The network converges to over 47.4\,dB PSNR, indicating that the distilled model closely approximates the iterative optimization results.
    }
    \label{fig:training_curve}
\end{figure}
\section{Additional Results}
\label{sec:supp:comparisons}

\subsection{Runtime Breakdown}
\label{sec:supp:runtime}

\Tbl{tab:runtime} reports the runtime breakdown of our distilled pipeline on an NVIDIA Jetson Xavier AGX.
Bilateral decomposition accounts for over 35\% of the total runtime.

\subsection{Distillation Training Convergence}
\label{sec:supp:training}

\Fig{fig:training_curve} shows the training convergence of our distilled network.
The validation loss decreases steadily over 60k iterations, and the network converges to over 47.4\,dB PSNR between the distilled output and the iterative optimization ground truth.
This high PSNR indicates that the distilled model closely approximates the results of the iterative optimization procedure.

\subsection{Progressive Optimization}
\label{sec:supp:progressive}

\begin{figure*}[t]
    \centering
    \includegraphics[width=0.6\textwidth, alt={Four paired examples comparing progressive optimization with direct optimization, which introduces high-frequency noise or exaggerated contrast.}]{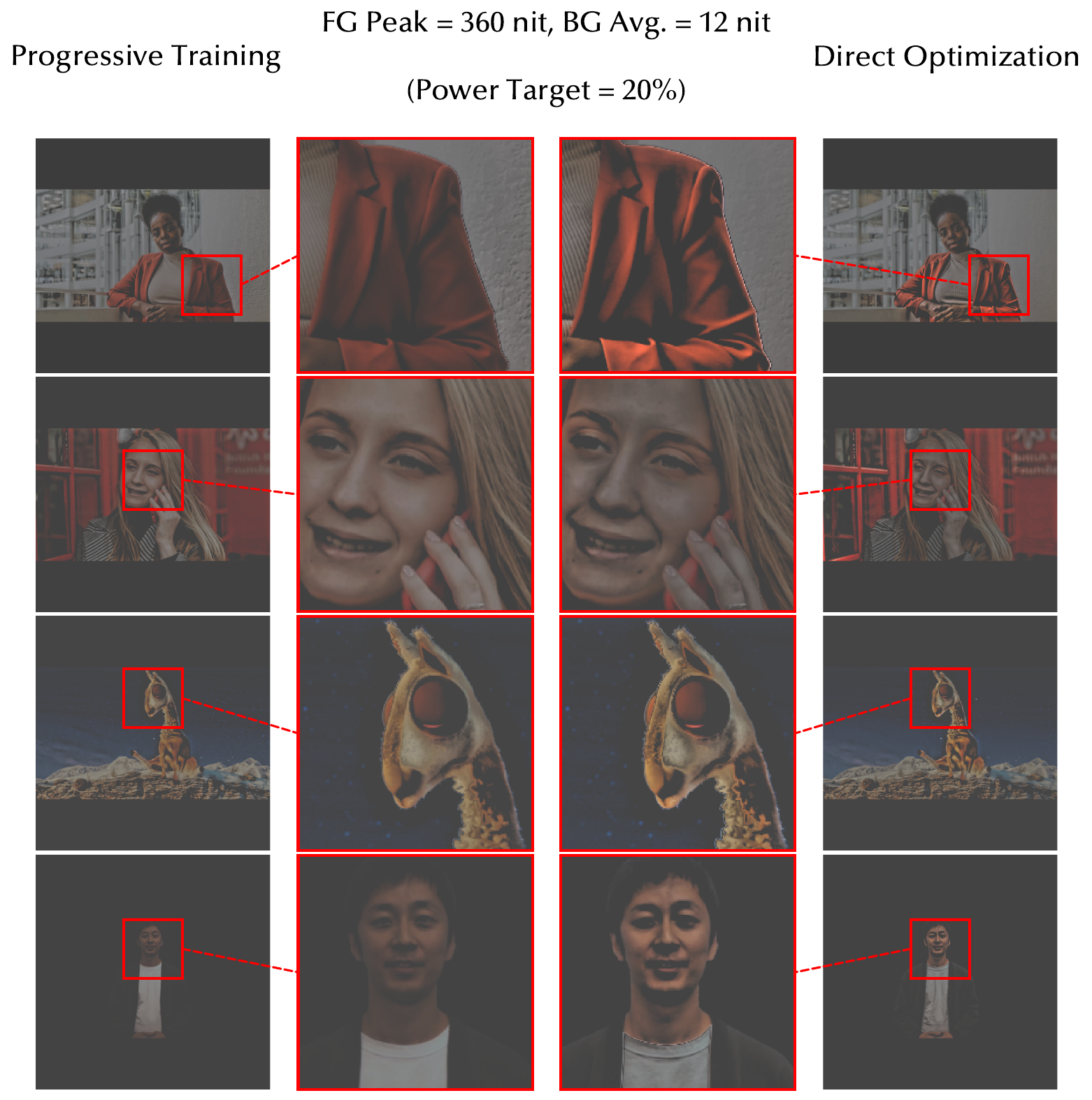}
    \caption{\textbf{Progressive optimization ablation.}
    Comparison between our progressive optimization (left in each pair) and direct optimization without progressive training (right in each pair) at 20\% power and FG:BG ratio of 30:1.
    Direct optimization produces artifacts, such as  high frequency noise or overly exaggerated contrast, due to unreliable gradients from the large appearance gap, while our progressive strategy maintains better results.
    }
    \label{fig:supp_progressive}
\end{figure*}

We provide additional comparisons between our progressive optimization strategy and direct optimization using the original full-power image as reference.
\Fig{fig:supp_progressive} shows four examples at a 20\% power target and FG:BG ratio of 30:1.
Without progressive optimization, the large appearance gap between the reference and target causes unreliable gradients, resulting in high-frequency noise artifacts (particularly visible at clothing boundaries and textured regions).
Our progressive strategy avoids these artifacts by ensuring that each optimization stage compares images with only small differences.

\subsection{Low-Power AR Perceptual Metrics}
\label{sec:supp:quant}

No metrics are designed specifically for low-power, additive displays.
With the subjective data (Fig.~\ref{fig:user_study}), we carry out a first-order analysis to determine how well widely used perceptual metrics can predict the relative orders of human preference.

We evaluate three popular perceptual metrics: ColorVideoVDP~\cite{mantiuk2024colorvideovdp}, MILO~\cite{cogalan2025milo}, and LPIPS~\cite{zhang2018unreasonable}.
The former accounts for absolute luminance, for which we provide physical nits values directly;
the latter operate on normalized $[0,1]$ intensity, so we normalize both the reference and test images to the same maximum value.
We use the original foreground image as the reference and the tone-mapped foreground composited with the background as the test image.
For the additive compositing, we use uniform backgrounds as suggested by Chapiro et al.~\cite{chapiro2024ar}.

\begin{figure*}[t]
    \centering
    \includegraphics[width=\linewidth, alt={Nine line charts comparing LowPowAR and four baselines with ColorVideoVDP, MILO, and LPIPS across power budgets and foreground-to-background luminance ratios of 3 to 1, 10 to 1, and 30 to 1.}]{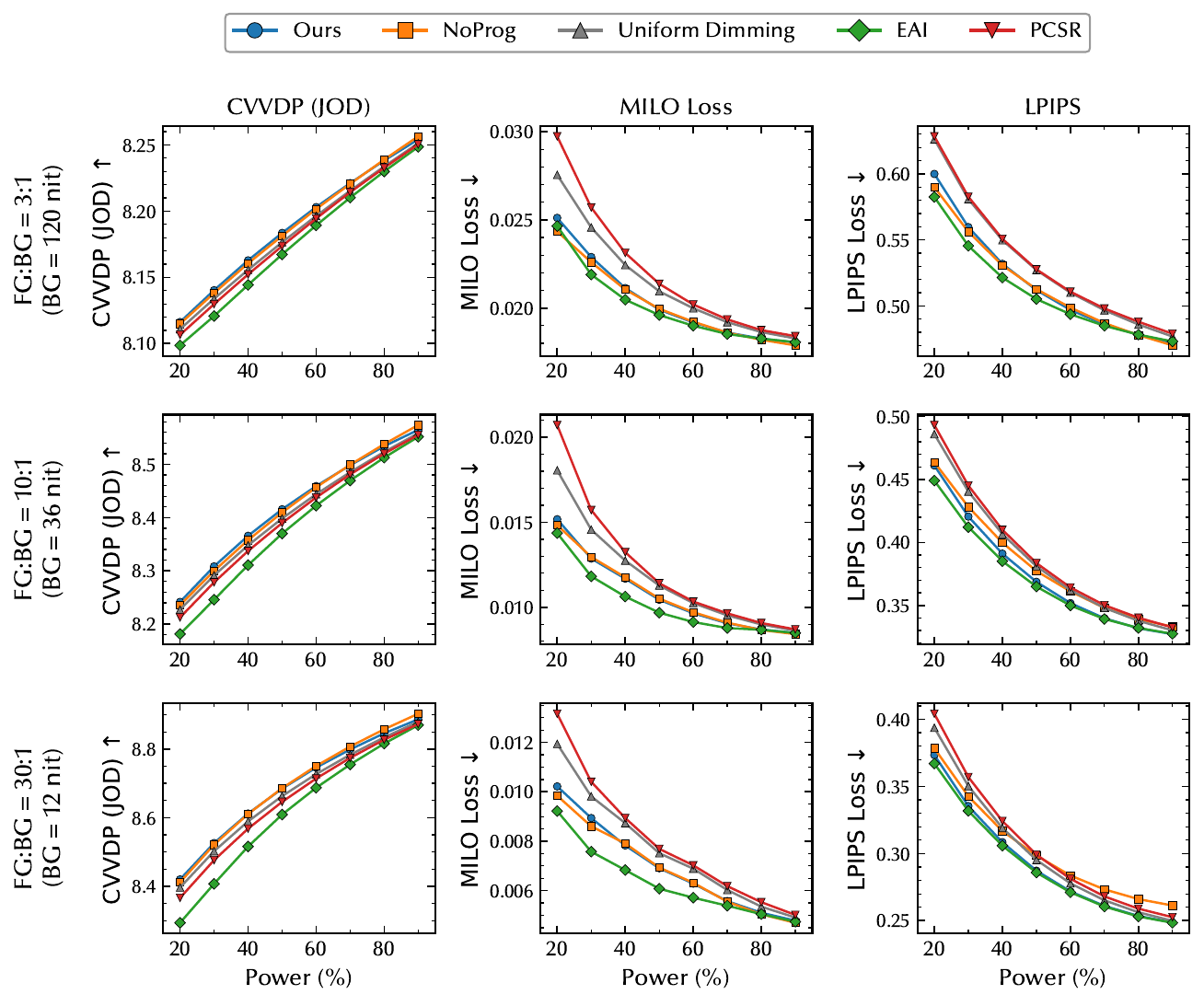}
    \caption{\textbf{Quality-power trade-off across three perceptual metrics and FG:BG ratios.}
    We compare all methods using ColorVideoVDP, MILO, and LPIPS at FG:BG ratios of 3:1, 10:1, and 30:1.
    ColorVideoVDP correctly ranks our method above baselines, consistent with user study results, while MILO and LPIPS favor \mode{EAI} despite its color artifacts.
    Foreground peak luminance is set to 360 nits.}
    \label{fig:supp_quant_ratio}
\end{figure*}

\Fig{fig:supp_quant_ratio} compares all methods, including the baselines used in our user study, across three perceptual metrics at three FG:BG ratios (3:1, 10:1, 30:1).
ColorVideoVDP correctly predicts that our method achieves the best quality-power trade-off across all settings.
Interestingly, MILO and LPIPS predict that \mode{EAI} achieves the highest quality (lowest loss), which contradicts the human data (Fig.~\ref{fig:user_study}).

That said, the absolute Just Objectionable Difference (JOD) values predicted by ColorVideoVDP should also be interpreted with caution: the predicted differences are less than 1 JOD between vastly different conditions (e.g., 3:1 vs.\ 30:1 FG:BG ratio, or 20\% vs.\ 90\% power), whereas subjectively the perceptual differences are much larger.
This finding is consistent with prior work~\cite{chapiro2024ar}, which shows that even ColorVideoVDP---the best-performing metric for AR content so far---does not correlate strongly with human judgments in AR scenarios.

\begin{figure*}[t]
    \centering
    \IfFileExists{supp/supp_figs/pq_metrics_all_ratios.pdf}{%
        \includegraphics[width=\linewidth, height=0.78\textheight, keepaspectratio, alt={PQ-encoded MILO and LPIPS results comparing LowPowAR and the evaluated baselines across foreground-to-background luminance ratios of 3 to 1, 10 to 1, and 30 to 1.}]{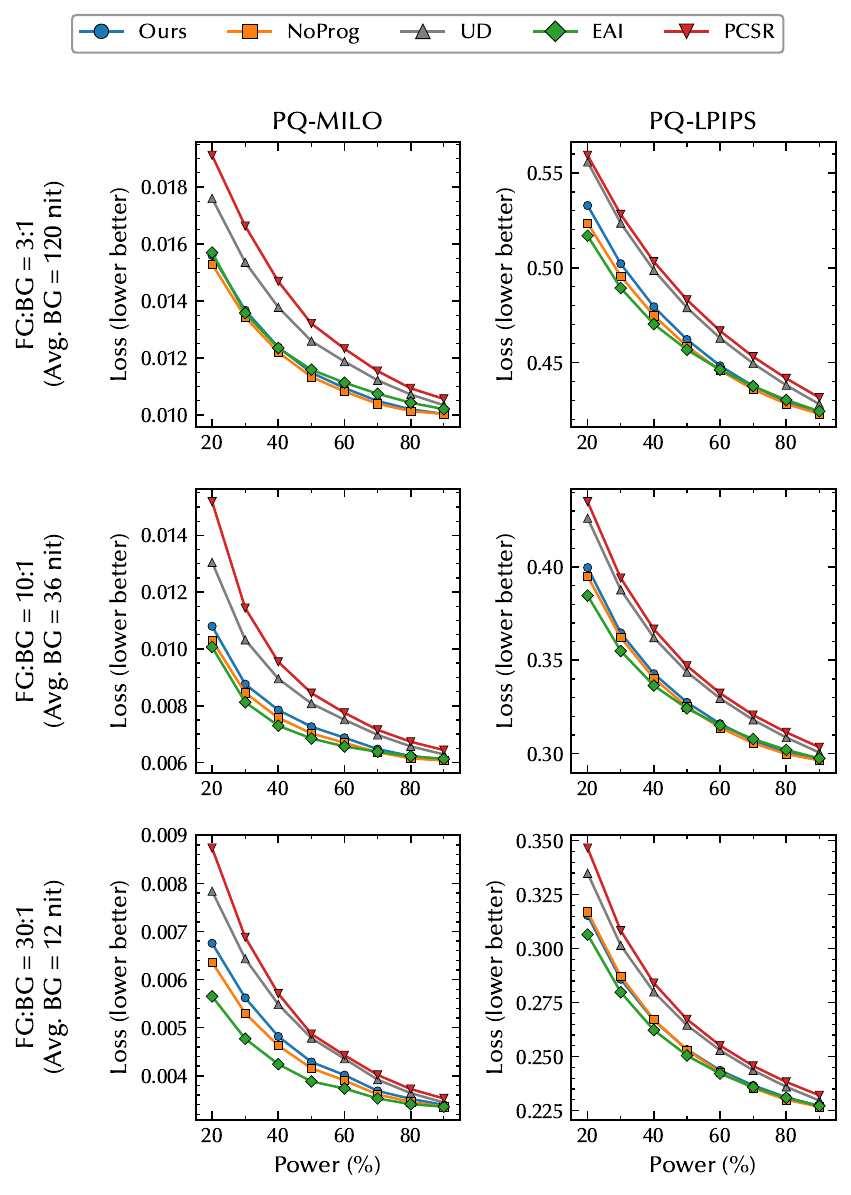}%
    }{%
        \fbox{\parbox[c][0.24\textheight][c]{0.94\linewidth}{%
            \centering
            Placeholder for\\[4pt]
            \texttt{supp\_figs/pq\_metrics\_all\_ratios.pdf}%
        }}%
    }
    \caption{\textbf{PQ-encoded perceptual-metric comparison.}
    PQ-MILO and PQ-LPIPS results across FG:BG ratios of 3:1, 10:1, and 30:1.}
    \label{fig:supp_pq_metrics}
\end{figure*}

We additionally test whether perceptual-quantizer input encoding improves agreement with the user study.
For this evaluation, we convert each image through the calibrated display model to absolute luminance and then apply the SMPTE ST~2084 PQ transfer function~\cite{smpte2014st2084} before evaluating MILO and LPIPS.
\Fig{fig:supp_pq_metrics} summarizes the PQ-encoded metric results.
PQ-LPIPS ranks \mode{EAI} first at every FG:BG ratio, while PQ-MILO ranks it second at 3:1 and first at 10:1 and 30:1.
Thus, even after PQ encoding, MILO and LPIPS remain inconsistent with the human data and generally continue to favor \mode{EAI}.

Our results point to the need to better understand perception in AR displays~\cite{zhang2021perceived, zhang2018color}, especially when the displayed stimuli are degraded by lowering the power budget.

\subsection{Direct Training vs.\ Distillation}
\label{sec:supp:distillation}

\begin{figure*}[t]
    \centering
    \includegraphics[width=0.56\linewidth, alt={Six examples comparing the reference, distilled LowPowAR output, and direct network training, with artifacts visible in the directly trained output.}]{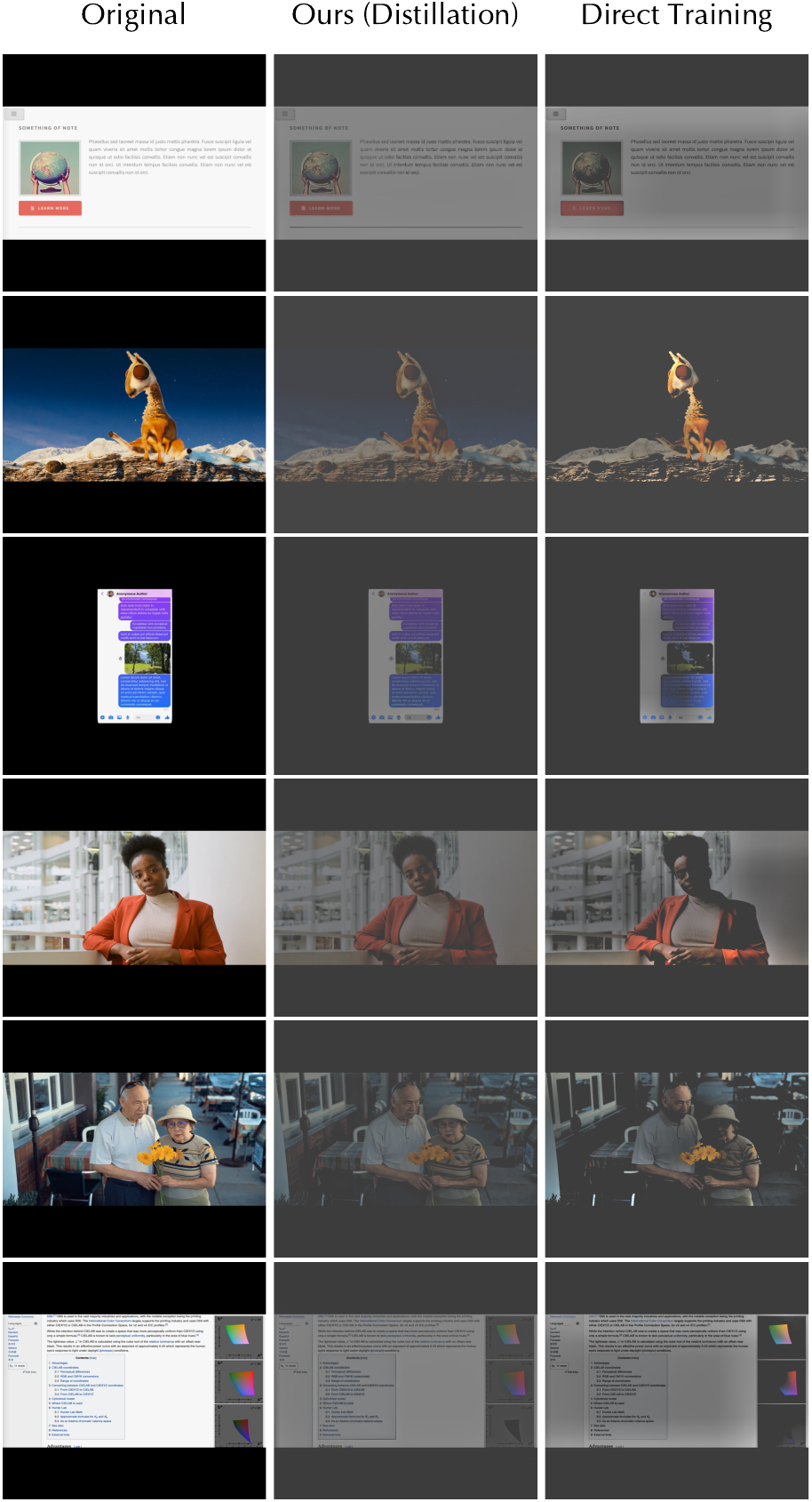}
    \caption{\textbf{Additional comparisons: Direct training vs.\ distillation.}
    Direct training with perceptual loss produces visible artifacts (right), while our distillation approach yields better results (middle).
    FG:BG\,=\,30:1, target power\,=\,20\%. All use foreground peak luminance of 360 nits.
    }
    \label{fig:supp_distillation}
\end{figure*}

We provide additional comparisons between direct training and our distillation approach.
As discussed in the main paper (\Sect{main-sec:exp:ablation:dt}), directly training the network with ColorVideoVDP and power loss tends to be unstable and converges to suboptimal solutions with visible artifacts.
\Fig{fig:supp_distillation} shows additional examples comparing our distillation approach (middle) versus direct training (right).
The results consistently show that direct training produces artifacts, while our distillation approach yields significantly better outputs across diverse foreground images.

\subsection{TMO Parameterization Comparison}
\label{sec:supp:tmo}

\begin{figure*}[t]
    \centering
    \includegraphics[width=0.7\textwidth, alt={Visual comparison of LowPowAR with gamma, Laplacian-pyramid, and Reinhard tone-mapping parameterizations under multiple power budgets.}]{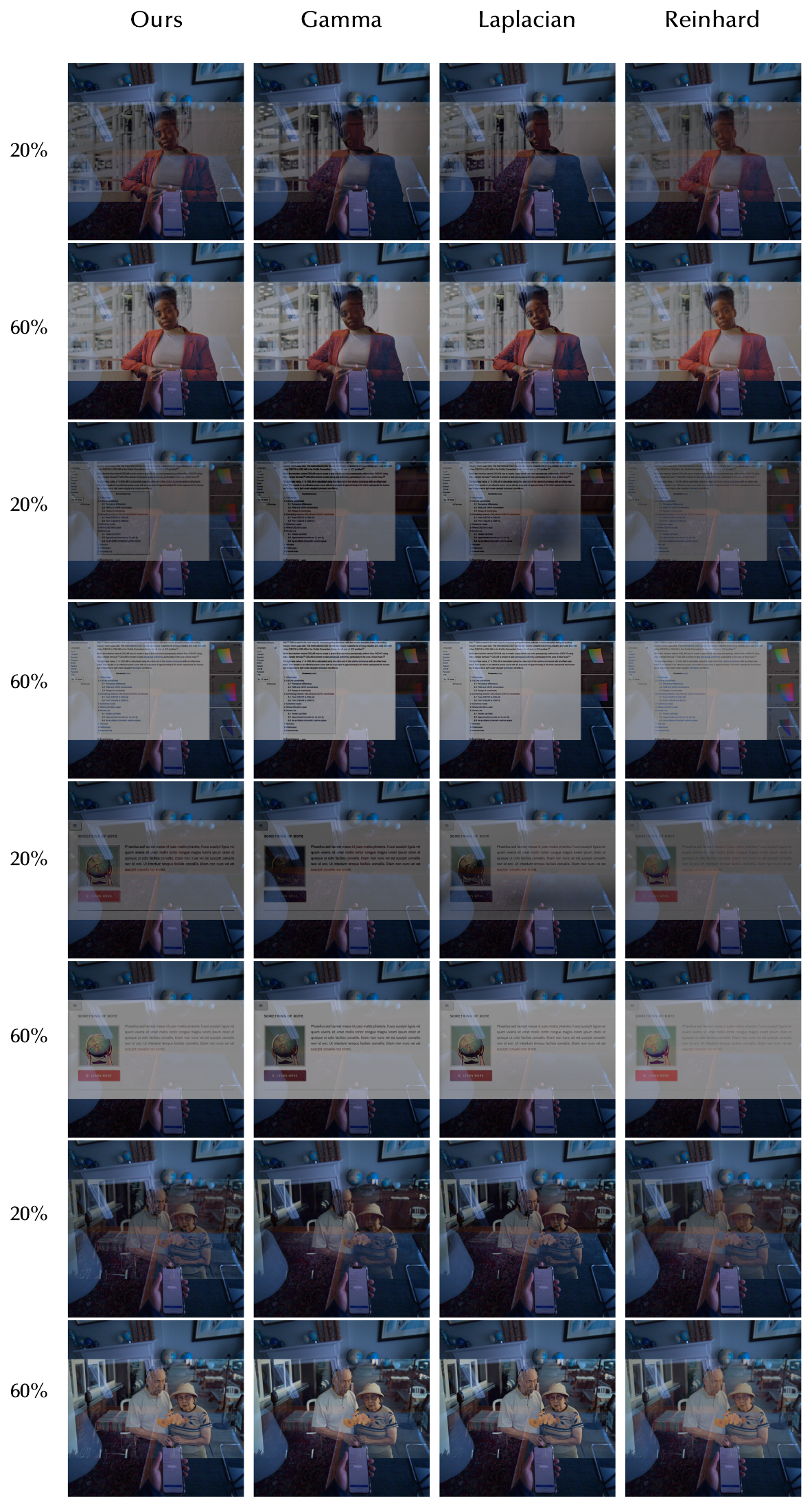}
    \caption{\textbf{Comparison of different TMO parameterizations.}
    Visual comparison of our bilateral decomposition approach against Gamma, Laplacian Pyramid, and Reinhard-based TMOs.
    Our method better preserves fine details and produces fewer artifacts across different power budgets.
    We use FG:BG = 30:1 here and set FG peak to 360 nits.}
    \label{fig:supp_tmo}
\end{figure*}

We compare our bilateral decomposition-based TMO against alternative tone mapping formulations.
\Fig{fig:supp_tmo} shows comparisons across multiple foreground content at different power budgets.

We evaluate the four TMO parameterizations described in \Sect{main-sec:exp:ablation:tmo}: Our Bilateral Decomposition, Gamma, Laplacian Pyramid, and Reinhard.
As shown in \Fig{fig:supp_tmo}, the Gamma and Laplacian Pyramid methods tend to over-enhance contrast, excessively darkening regions that are already dim in the original image, leading to visible artifacts.
Reinhard's photographic tone mapping avoids this over-darkening issue.
However, our bilateral decomposition approach better preserves high-frequency details (e.g., text, textures) by explicitly separating base and detail layers, allowing independent control over local luminance adaptation and detail enhancement.

\subsection{Uniform vs.\ Precise Background}
\label{sec:supp:uniform_vs_precise}

\begin{figure*}[t]
    \centering
    \includegraphics[width=0.95\textwidth, alt={Three foreground-background pairs comparing tone mappings optimized with an average uniform background and with the precise spatial background across several conditions.}]{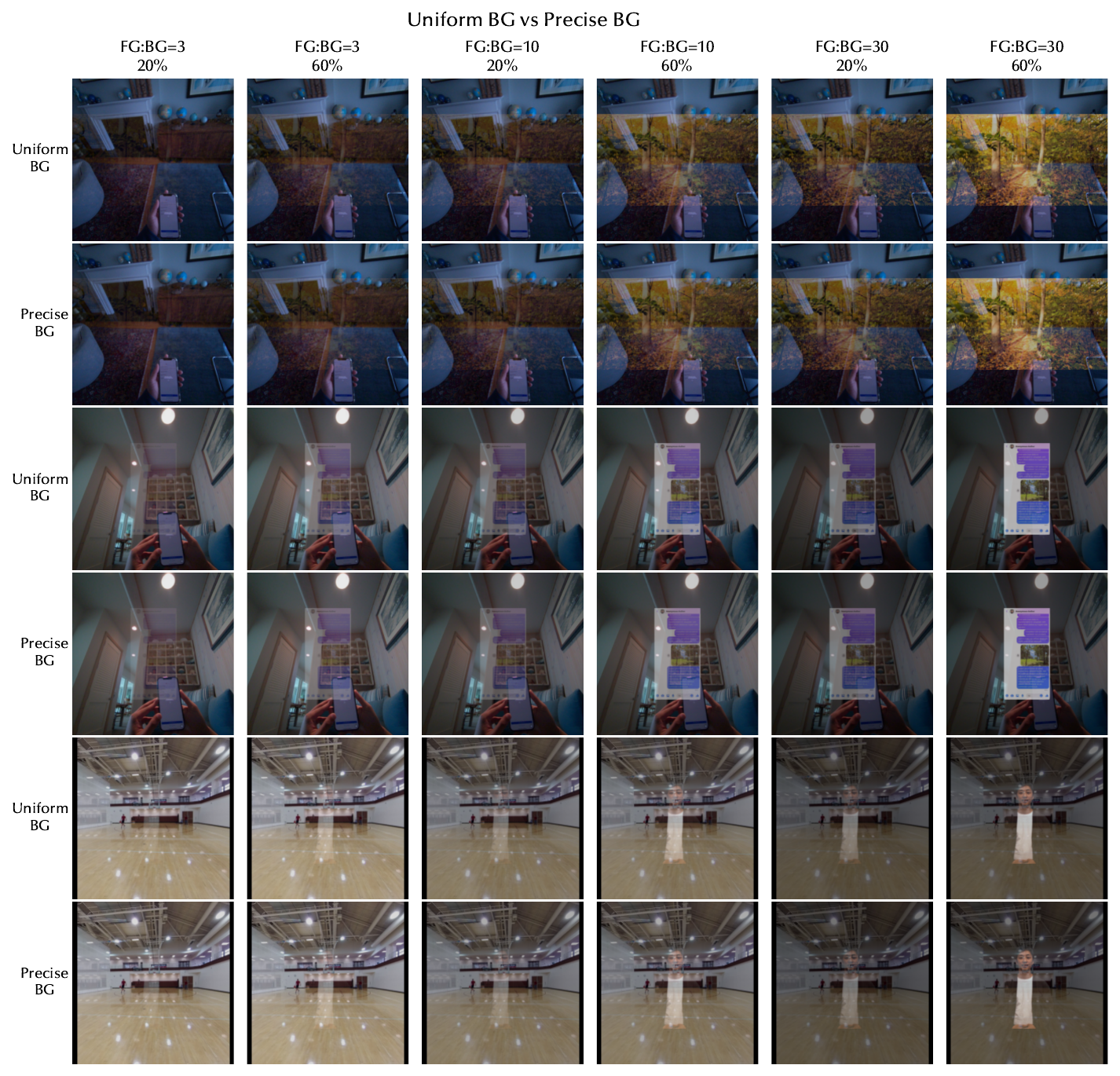}
    \caption{\textbf{Uniform vs.\ precise background comparison (Part 2).}
    Additional three foreground-background pairs comparing uniform background (odd rows) versus precise background (even rows) optimization.
    The consistent similarity across pairs confirms that average luminance is sufficient for optimization.}
    \label{fig:supp_uniform_precise_2}
\end{figure*}

We compare our method using uniform background representation versus precise background representation for optimization.
As discussed in the main paper, we use only the average background luminance to derive the display mapping operators.
 \Fig{fig:supp_uniform_precise_2} shows additional comparisons across six foreground-background pairs, three FG:BG ratios (3, 10, 30), and two power budgets (20\%, 60\%).
The results demonstrate that uniform background optimization produces comparable or sometimes superior results to precise background optimization, validating our design choice.

\subsection{Bilateral Decomposition}
\label{sec:supp:decomposition}

\begin{figure*}[t]
    \centering
    \includegraphics[width=0.95\textwidth, alt={Three foreground-background pairs comparing bilateral base-detail tone mapping with direct piecewise tone mapping, showing better detail preservation with decomposition.}]{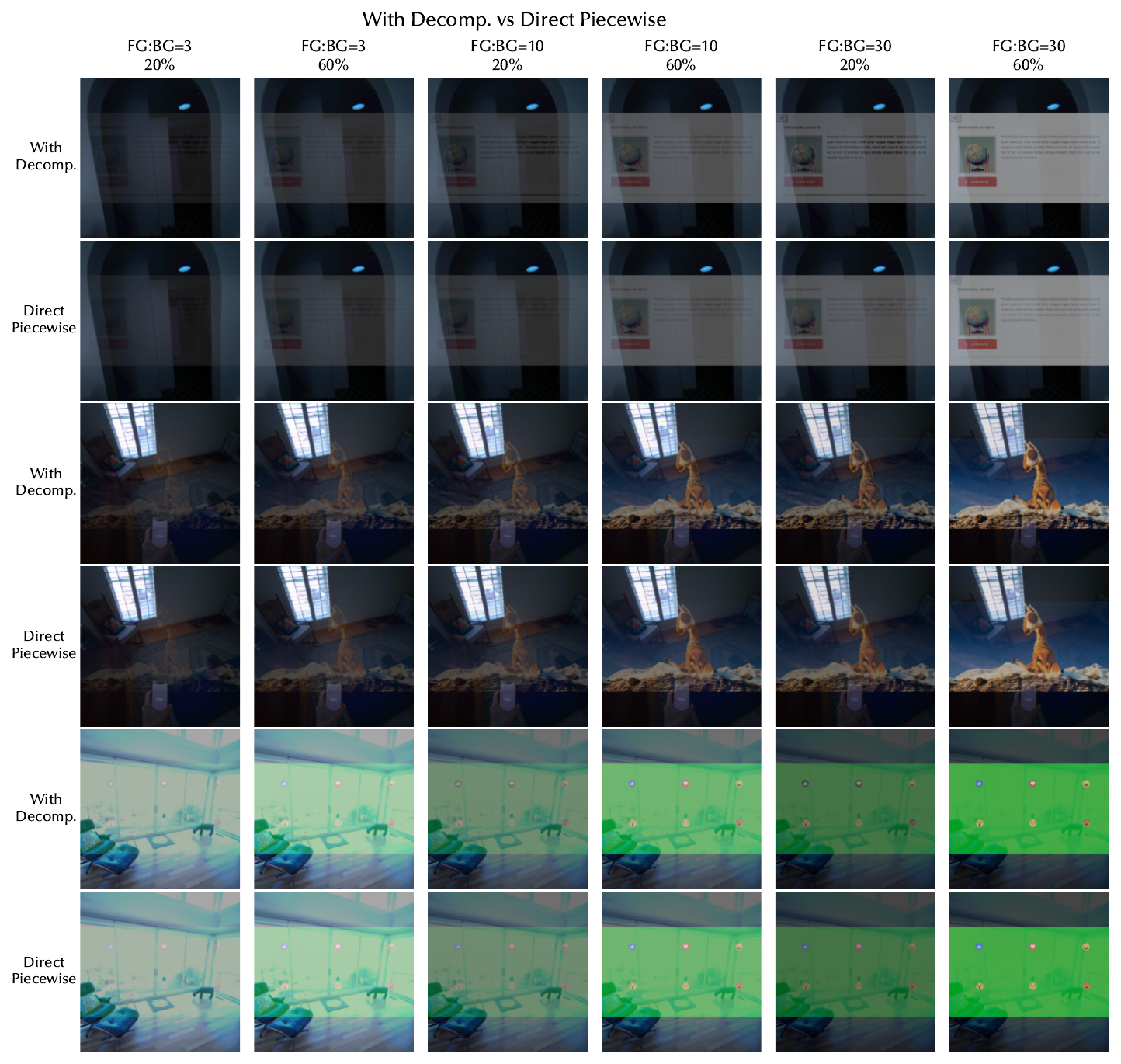}
    \caption{\textbf{Bilateral decomposition ablation (Part 1).}
    Our method with bilateral decomposition (odd rows) versus direct piecewise tone mapping (even rows) for three foreground-background pairs.
    Bilateral decomposition better preserves high-frequency details across all conditions.}
    \label{fig:supp_decomp_1}
\end{figure*}

We compare our method with bilateral decomposition against direct piecewise tone mapping without decomposition.
\Fig{fig:supp_decomp_1}  shows additional comparisons across six foreground-background pairs.
Our method with bilateral decomposition consistently preserves high-frequency details better than direct piecewise mapping, particularly visible in text regions and fine textures.

\subsection{Limitation: Small Text Artifacts}
\label{sec:supp:limitation}

\begin{figure}[t]
    \centering
    \includegraphics[width=\linewidth, alt={Examples showing noise and reduced legibility on small dark text over a white foreground canvas after tone mapping.}]{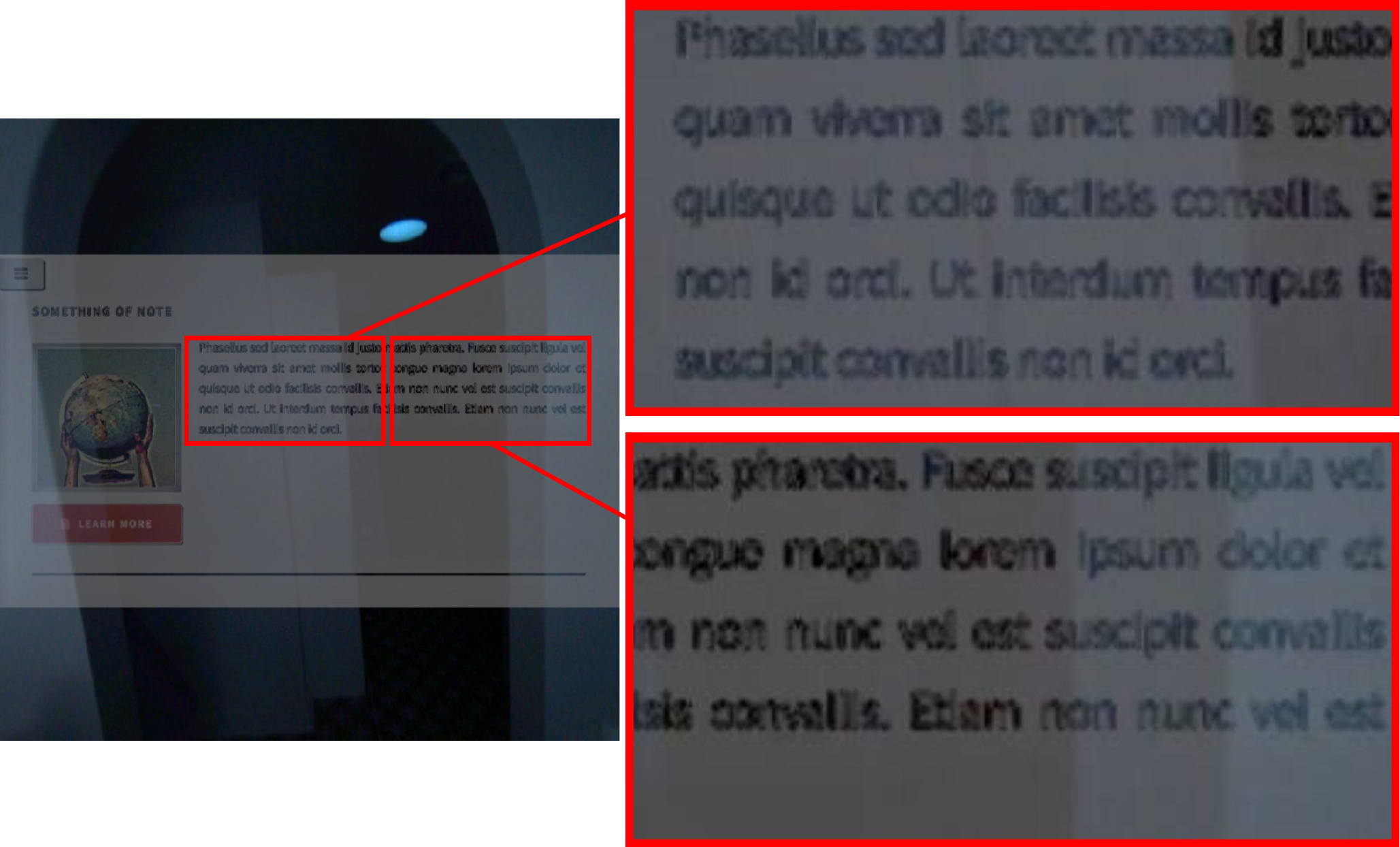}
    \caption{\textbf{Limitation: small text artifacts.}
    Our tone mapping can introduce noise on small dark text over a white canvas, reducing legibility.}
    \label{fig:supp_limitation}
\end{figure}

\Fig{fig:supp_limitation} shows examples of the text legibility artifact discussed in the main paper (\Sect{main-sec:exp:limitation}).
When the foreground contains small dark text on a white canvas, our tone mapping can introduce noise that reduces readability.
This issue does not arise for white text on a black canvas (see small white text in supplementary video \texttt{tiktok\_web\_browse.mp4} in \texttt{video\_tmo\_examples.zip} for example).

\subsection{More Visual Comparisons against Baselines}
\label{sec:supp:baselines}

\begin{figure*}[p]
    \centering
    \includegraphics[width=0.9\textwidth, alt={Visual comparison of LowPowAR, NoProg, uniform dimming, EAI, and PCSR for a Caminandes foreground and Aria location 2 background.}]{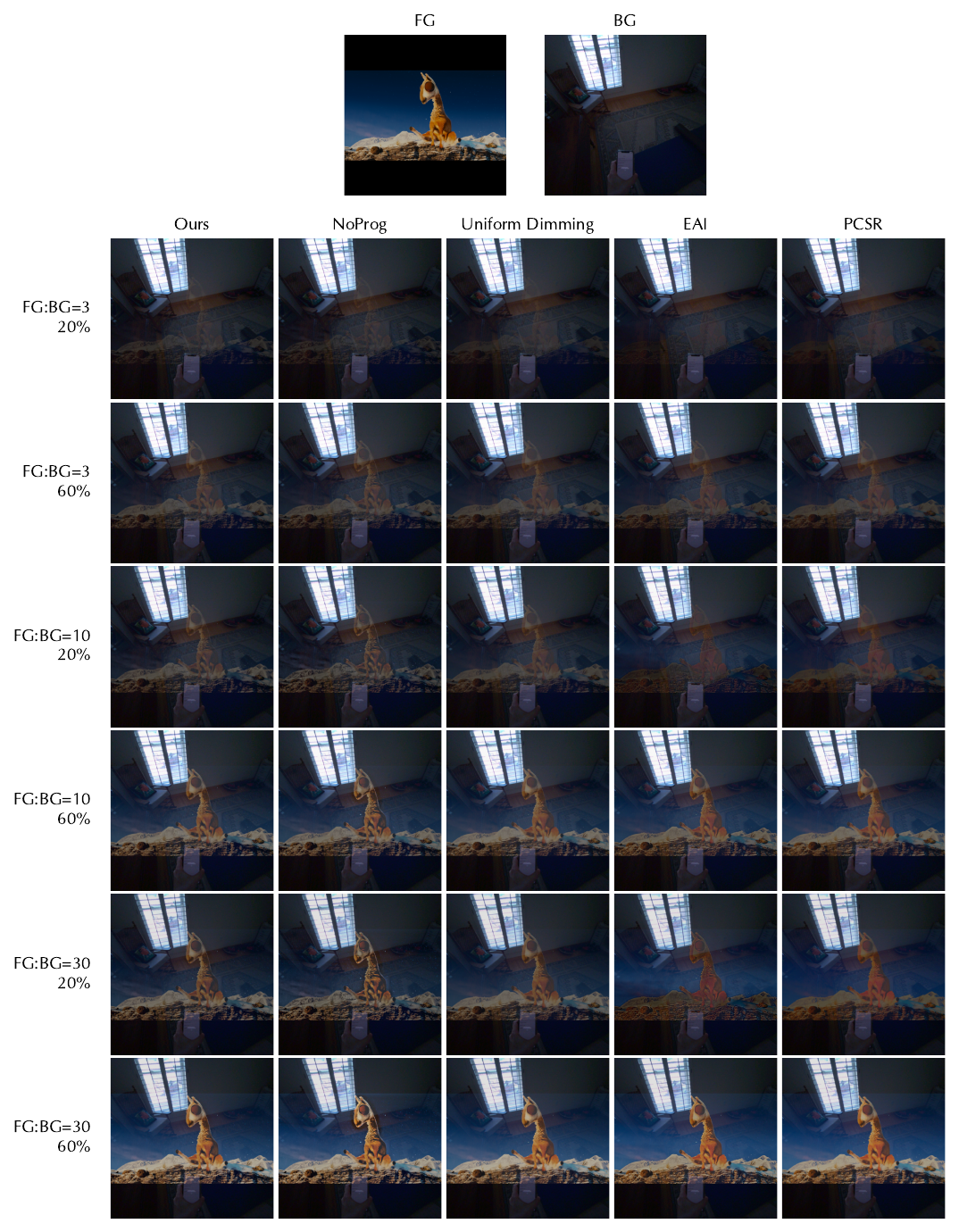}
    \caption{Visual comparison: Caminandes foreground with Aria location 2 background.}
    \label{fig:supp_blog_aria2}
\end{figure*}

\begin{figure*}[p]
    \centering
    \includegraphics[width=0.9\textwidth, alt={Visual comparison of the five evaluated methods for a messaging foreground and Aria location 5 background.}]{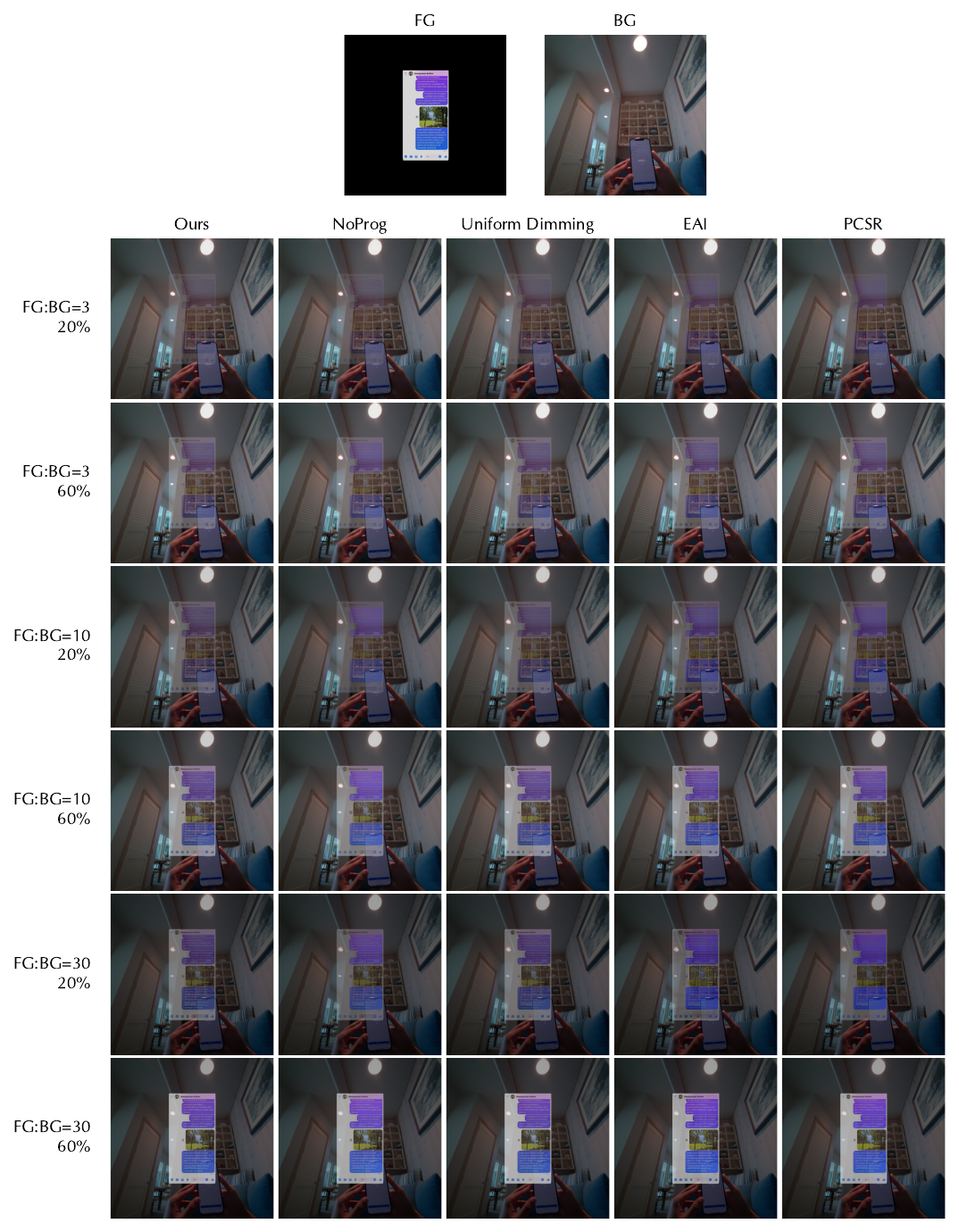}
    \caption{Visual comparison: Messaging foreground with Aria location 5 background.}
    \label{fig:supp_messaging_aria5}
\end{figure*}

\begin{figure*}[p]
    \centering
    \includegraphics[width=0.9\textwidth, alt={Visual comparison of the five evaluated methods for a bonfire foreground and PeaPod Sculpture background.}]{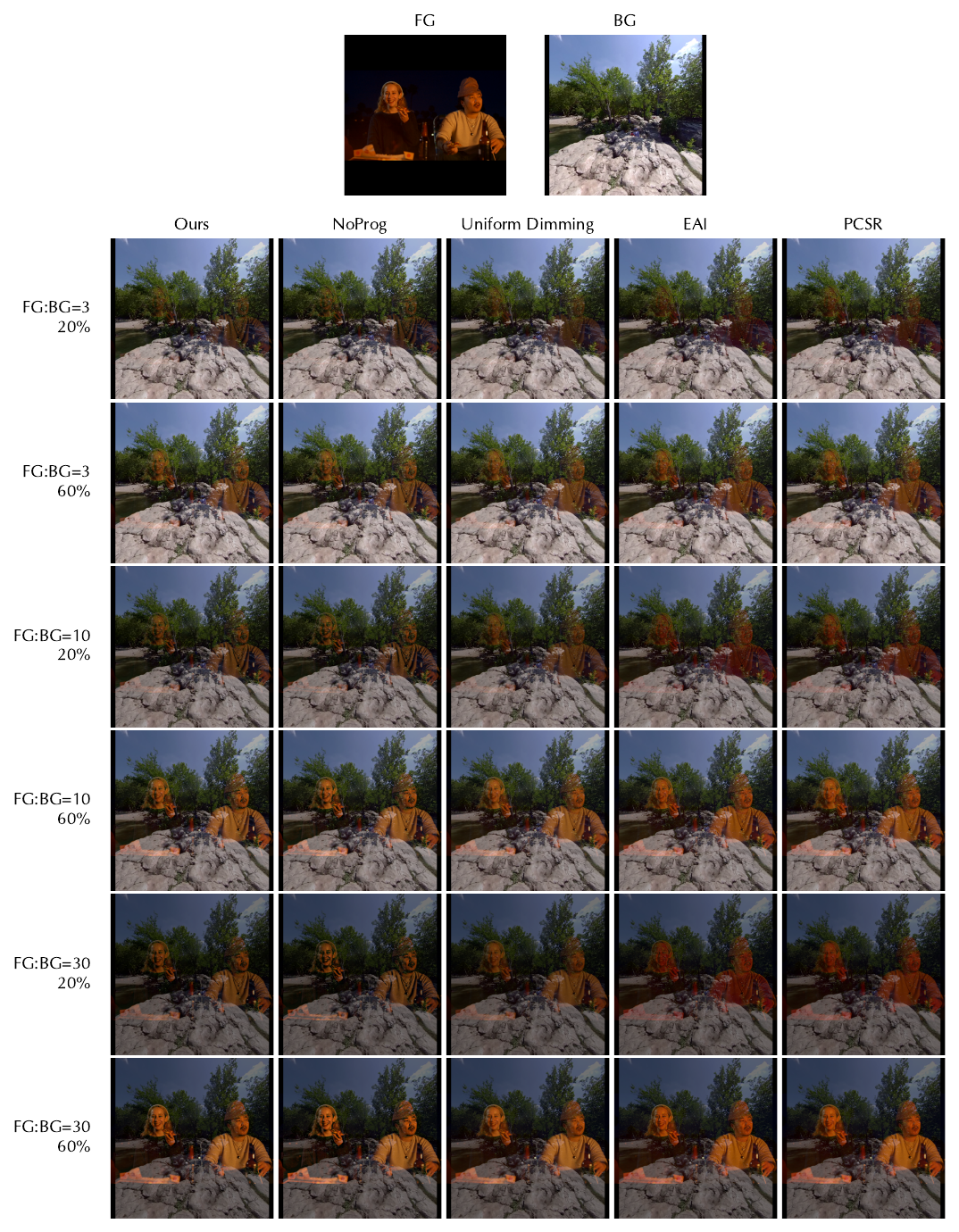}
    \caption{Visual comparison: Bonfire foreground with PeaPod Sculpture background.}
    \label{fig:supp_bonfire_sculpture}
\end{figure*}

\begin{figure*}[p]
    \centering
    \includegraphics[width=0.9\textwidth, alt={Visual comparison of the five evaluated methods for a business foreground and PeaPod Street background.}]{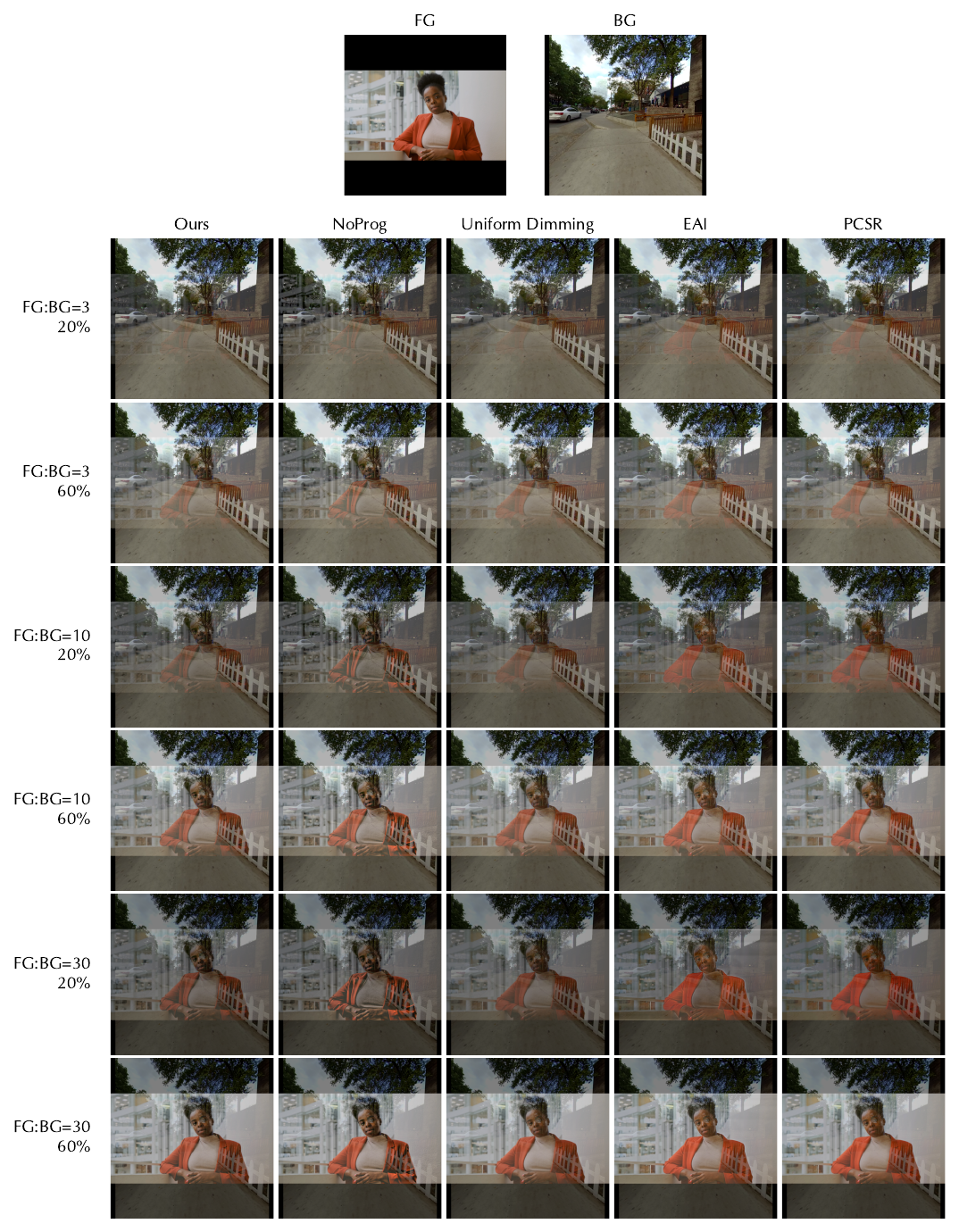}
    \caption{Visual comparison: Business foreground with PeaPod Street background.}
    \label{fig:supp_business_street}
\end{figure*}

\begin{figure*}[p]
    \centering
    \includegraphics[width=0.9\textwidth, alt={Visual comparison of the five evaluated methods for an icons foreground and dead-leaves background.}]{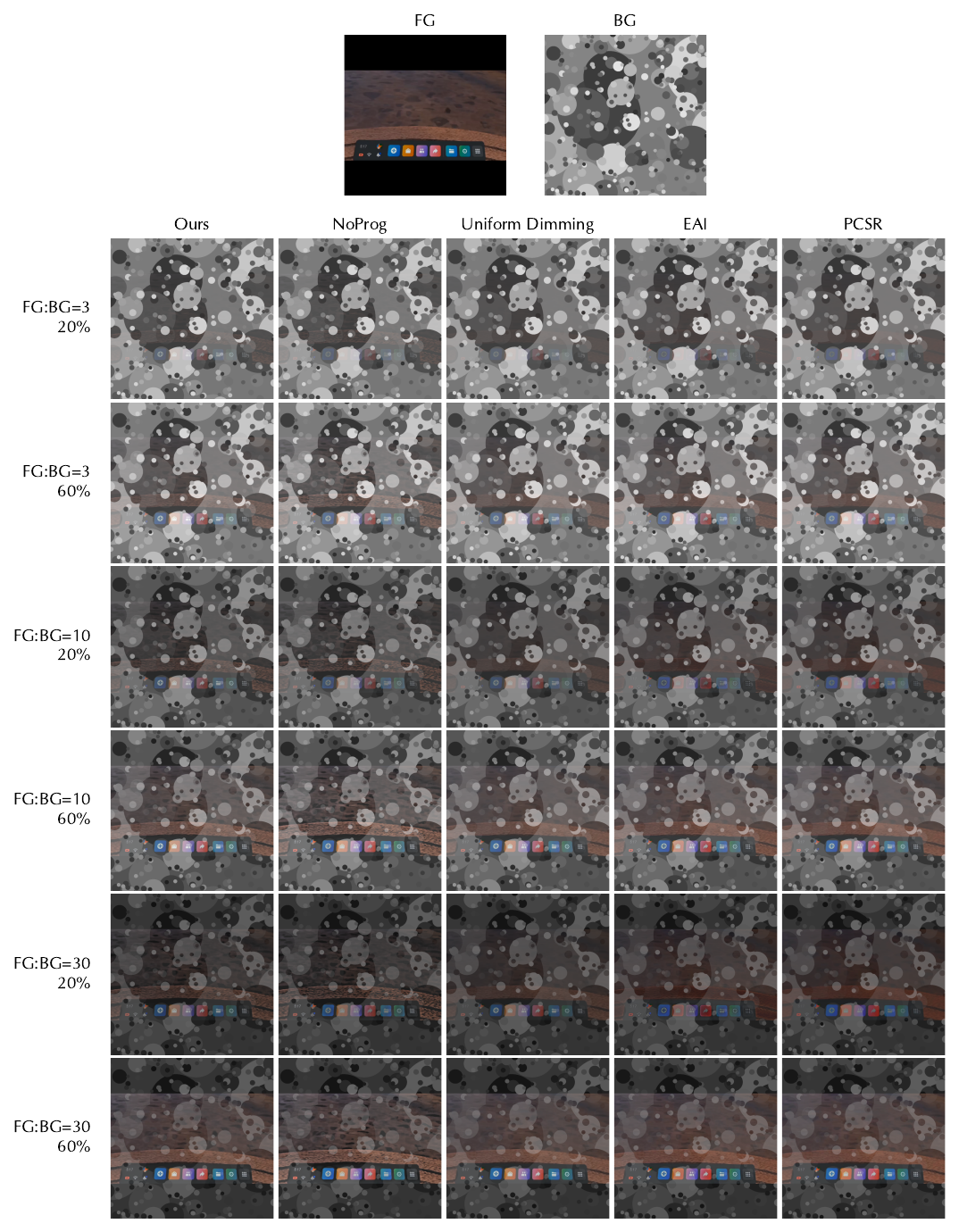}
    \caption{Visual comparison: Icons foreground with dead leaves background.}
    \label{fig:supp_icons_deadleaves}
\end{figure*}

\begin{figure*}[p]
    \centering
    \includegraphics[width=0.9\textwidth, alt={Visual comparison of the five evaluated methods for a panel foreground and flat uniform background.}]{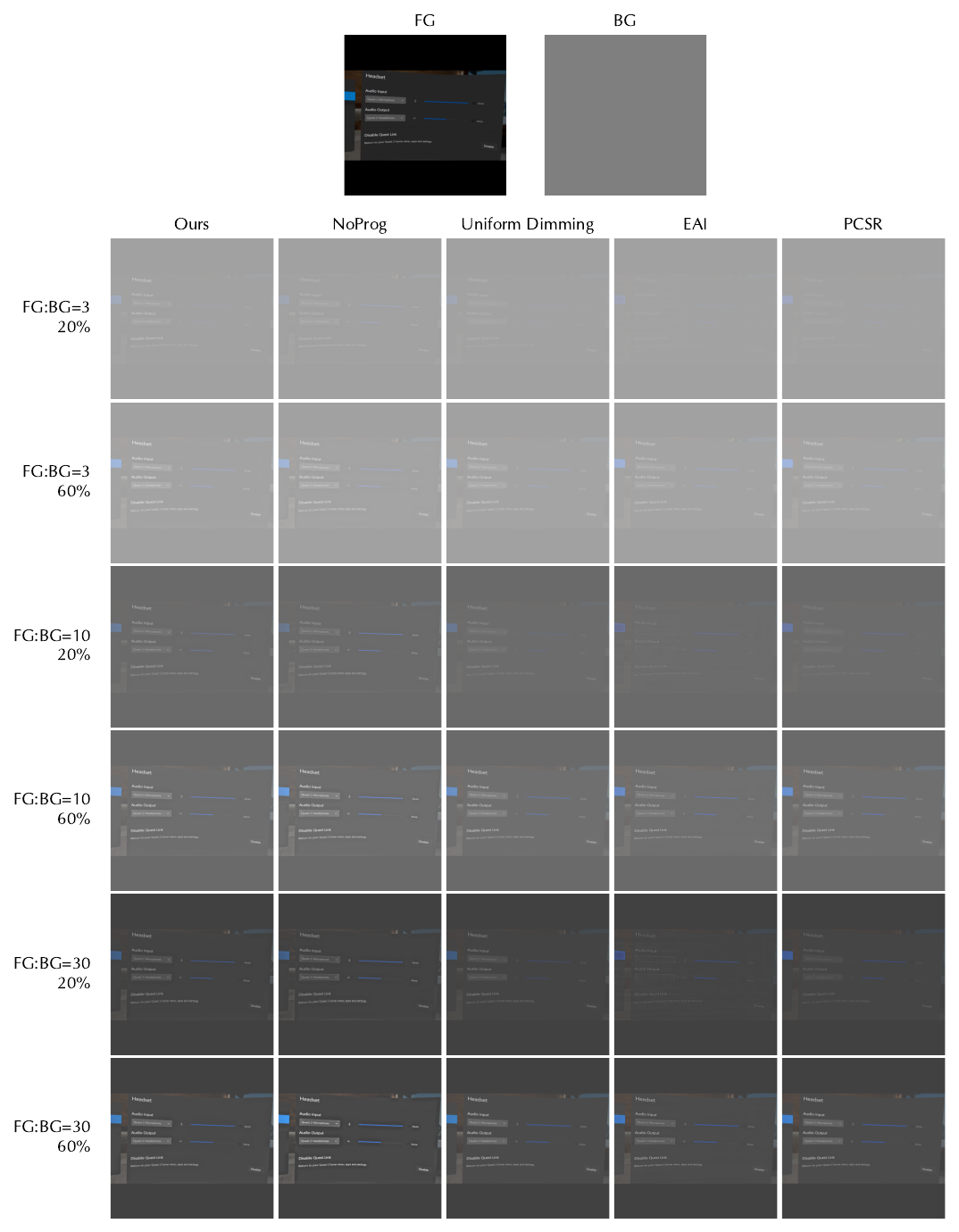}
    \caption{Visual comparison: Panel foreground with flat (uniform) background.}
    \label{fig:supp_panel_flat}
\end{figure*}

\begin{figure*}[p]
    \centering
    \includegraphics[width=0.9\textwidth, alt={Visual comparison of the five evaluated methods for a phone foreground and pink-noise background.}]{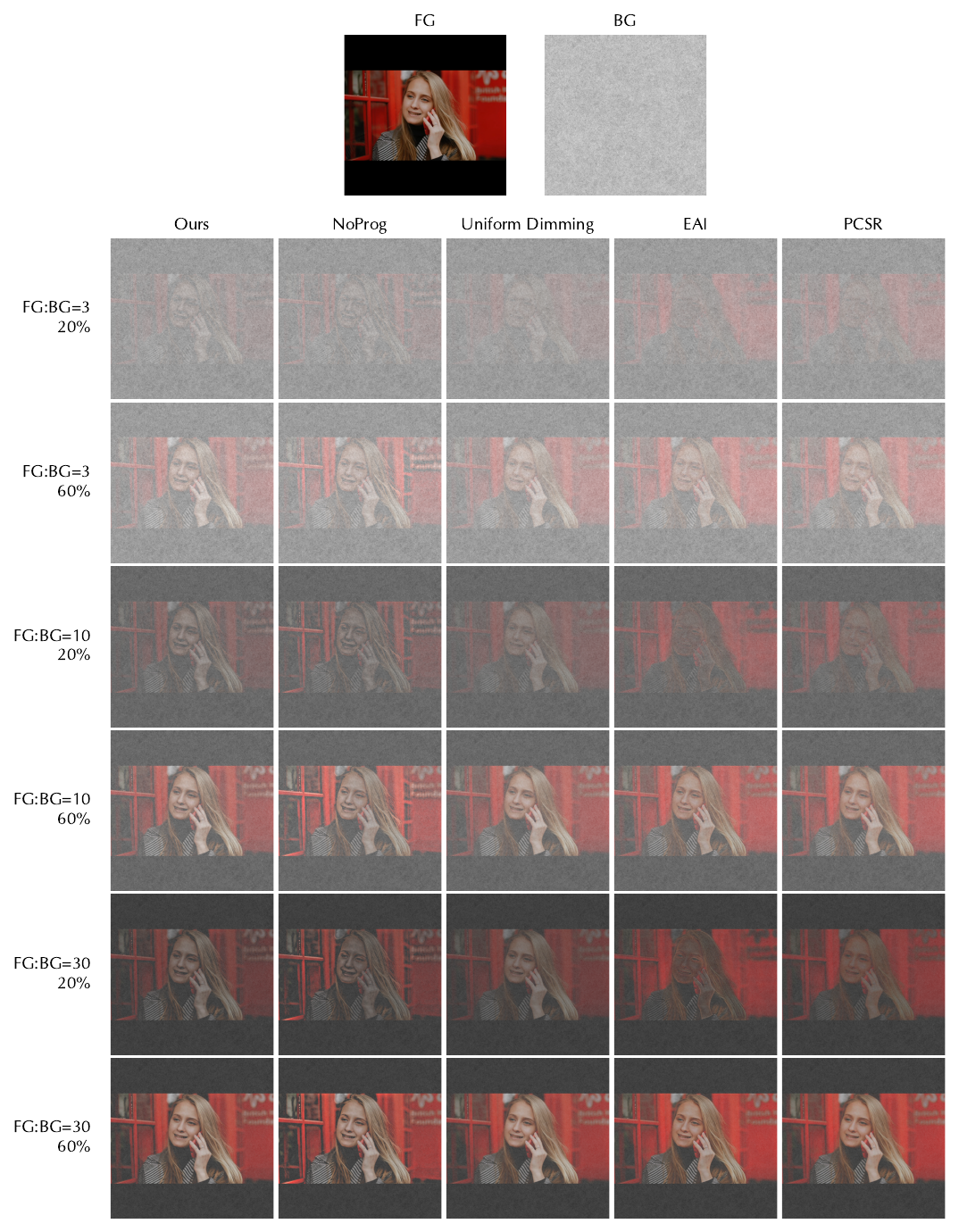}
    \caption{Visual comparison: Phone foreground with pink noise background.}
    \label{fig:supp_phone_deadleaves}
\end{figure*}

We provide additional visual comparisons against baselines across diverse foreground-background pairs (\Fig{fig:supp_blog_aria2} -- \Fig{fig:supp_phone_deadleaves}).

\clearpage

\clearpage

\clearpage

\clearpage
\clearpage

\clearpage

\clearpage

\clearpage

\clearpage

\clearpage

\clearpage

\clearpage

\clearpage

\clearpage

\end{document}